\documentclass[9pt,twocolumn,a4paper]{article}

\usepackage{amsmath,graphicx,color,amssymb,caption} 
\usepackage{hyperref}
\hypersetup{linktocpage}

\usepackage[text={17.7truecm,24.4truecm},hcentering]{geometry}

\author{S. Fratini$^*$, D. Mayou, and S. Ciuchi}

\title{The Transient Localization Scenario for 
Charge Transport in Crystalline Organic Materials.}

\begin{document}

\maketitle
\begin{abstract}\small \bf
Charge transport in crystalline organic
semiconductors is intrinsically limited by the presence of large thermal
molecular motions, which are a direct consequence of the weak van der Waals
inter-molecular interactions. These lead to an original regime of transport called
\textit{transient localization}, sharing features of both localized and itinerant
electron systems. After a brief review of experimental observations that pose a
challenge to the theory, we concentrate on a commonly studied model which
describes the interaction of the charge carriers with inter-molecular
vibrations. We present different theoretical approaches that have been applied
to the problem in the past, and then turn to more modern
approaches that are able to capture the key microscopic phenomenon at the origin
of the puzzling experimental observations, i.e. the quantum localization of the
electronic wavefuntion at timescales shorter than the typical molecular motions.
We describe in particular a relaxation time approximation which clarifies how
the transient localization due to dynamical molecular motions relates to
the Anderson localization realized for static disorder, and 
allows us to devise strategies to improve the mobility of actual compounds.
The relevance of the transient localization scenario to other classes of systems is
briefly discussed.
\end{abstract}

\hrulefill

\noindent {\small
[*] Dr. S. Fratini, Dr. D. Mayou\\
Institut N\'{e}el, CNRS and Universit\'e Grenoble Alpes, F-38042 Grenoble, France\\
E-mail: simone.fratini@neel.cnrs.fr

\smallskip
\noindent Prof. S. Ciuchi\\
University of L'Aquila, Department of Physical and Chemical Sciences, Via Vetoio, L'Aquila, Italy;
CNR-ISC, Via dei Taurini, Rome, Italy; 
CNISM, Udr L'Aquila, Italy
}

\tableofcontents



\section{Introduction}

Molecular organic  semiconductors, i.e.  solids composed of small organic molecules,
have gained a rising interest in the past years because of their implementation as active
semiconductor layers in electronic and opto-electronic devices. These range from  
field-effect transistors to lightning devices and displays,  photovoltaic cells and novel spintronic devices. 
In parallel with such applied developments, remarkable advances have been made in understanding  
the electronic properties of  organic semiconductors, triggered by the great improvements in sample and device fabrication.
While in the past, the physical properties in these compounds were often masked by impurity effects or
structural inhomogeneities, it has now become possible to investigate high-quality crystals
via a broad panel of experimental techniques, 
giving access to the intrinsic  properties of the electronic charge carriers 
and their correlation with the microscopic material parameters. 

The quantity which has concentrated most efforts of both applied 
and fundamental studies is the charge carrier mobility $\mu$. This is a key material parameter because it 
influences the performances of actual semiconductor devices, such as
 the switching rate in transistors or the efficiency of energy transfer processes in photovoltaics.  
 \cite{BredasChemRev04,BredasACR2009}
At present, mobilities in excess of $10 cm^2/Vs$ can be achieved in organic
field-effect transistors (OFETs) based on crystalline organic solids, \cite{MinderAdvMat2011} 
and comparable values are also observed in OFETs with ordered films. \cite{SirringhausNatMat2010} 
Such values are much higher than those characterizing 
amorphous organic films ($10^{-4}-10^{-5} cm^2/Vs$) 
but remain orders of magnitude lower than those found in wide-band semiconductors,  
reaching  values higher than $10^7$
in inorganic semiconductor heterostructures, 
$10^6$ in graphene,  $10^3$ in crystalline silicon and several $10^2$ 
in transition-metal dichalcogenides.

This intrinsically
low mobility indicates extremely short electronic mean-free paths --- on the order
of the intermolecular distances --- 
leading to a breakdown of the basic assumptions underlying  band transport theory. 
It is currently believed that the
mobility in crystalline organic semiconductors, at least within the
technologically relevant  regime around room temperature, 
is intrinsically limited by the presence of large thermal molecular motions, which
are a direct consequence of the weak van der Waals intermolecular interactions. 
Deviations from the perfect crystalline arrangement, and thus from a
periodic Bloch-state, act as a dynamical source of disorder on the already
narrow electronic bands arising from the $\pi$-intermolecular overlaps. These induce
a localization of the electronic wavefunctions, which survives up to the typical time scales of the
inter-molecular vibrations. This phenomenon, that is not described
by the semi-classical Boltzmann theory of electron-phonon scattering,  
results in an original transport mechanism that was termed \textit{transient localization}. 
It is our purpose here to review the recent theoretical advances 
and experimental successes of the transient localization scenario for charge transport in 
crystalline organic solids.


\paragraph{Organic semiconductors: between the solid state and molecular pictures. --}
The order-of-magnitudes 
difference in the mobility of even the best organic semiconductors as compared to their wide-band counterparts
testifies that charge transport in these materials is not governed by the same  microscopic mechanisms.
In the early days of organic semiconductor research
more than 50 years ago, several alternative approaches were proposed. 
Attempts to describe organic semiconductors were made either extending the  realm of semi-classical Boltzmann theory beyond the standard treatments 
which apply to inorganic semiconductors,  
or using radically different theoretical concepts to start with,  such as Marcus theory and Holstein small
polaron theory, i.e. focusing on the molecular rather than the "band" character of these compounds. 
\cite{DukeSchein80} However, none of these 
has provided a truly satisfactory and consistent
description of the charge dynamics. 
In the years following these early works, researchers 
concentrated their efforts in extending such original concepts and
including more realistic material details, 
determining the microscopic parameters from  \textit{ab initio} methods 
 with ever increasing accuracy,
which was made possible by the great improvements in the numerical calculation capabilities.
It is however 
fair to say that no real theoretical breakthrough  emerged from these studies:
the well-known Marcus formula for hopping transport, for instance, still remains widely used today
to evaluate the mobility in fully \textit{ab initio} treatments of specific compounds, 
even though its very assumptions are violated in high-mobility organic semiconductors.

 From a theoretical point of view, the difficulty in addressing charge transport in organic materials
comes from the fact that several microscopic interactions are at work, 
whose characteristic energy scales are all of comparable magnitude, thus
preventing the applicability of standard limiting treatments.
The typical values of the intermolecular  transfer integrals, 
which determine the bandwidth of extended electronic states moving through the solid,
are in the range $J\sim 10-100meV$.
This is comparable to the  energy gained upon deformation of the individual  molecules to accommodate
excess charge carriers (i.e. the polaron energy or relaxation energy), 
$E_P\sim 50-200meV$,  \cite{DevosPRB1998,CoropceanuPRL2002,BlaseEphPRB2011,GirlandoPRB10}
which instead favors localization of the carriers on individual molecules. 
This "intermediate coupling" situation prevents in principle the use of both  Marcus theory,
which is  only applicable  when the inter-molecular transfer energy $J$ is the smallest energy 
scale in the problem, and standard band theory, which instead requires it to be the largest: neither of these limits is fulfilled  organic semiconductors materials.
Another important parameter which influences the carrier dynamics is the zero-point energy of molecular 
vibrations. In the case of the intra-molecular modes, these also
lie in the range $\hbar\Omega_0 \simeq 100- 200meV$. 
The carriers also interact with lower energy inter-molecular modes, 
in the range $\hbar \omega_0\lesssim 10 meV$.
\cite{GirlandoPRB10,SinovaPRL2001,TroisiPRL06}
Finally, the thermal energy  $k_BT\sim 25 meV$ at the ambient 
conditions relevant for technological applications is also of the same order of magnitude. 

The absence of an identified small parameter makes it difficult to find a proper starting point
on which to apply perturbation expansions, which  
is  often 
at the origin of long-standing and unsolved 
problems in physics. In this respect, organic charge transport is no exception to the rule.

\paragraph{The emergence  of an alternative paradigm. --}
In this theoretical no man's land, an alternative scenario that was proposed early on 
to describe charge transport in organic semiconductors 
(see Ref. \cite{DukeSchein80} for an early review)
 has been gaining strong support  in the last decade. 
It starts from the observation that, because the intermolecular Van der Waals forces 
which hold organic solids together are weak, thermal molecular motions in these materials can be very large.
Such motions are slow due to both the weak restoring forces and the large molecular masses, 
and provide the electrons at each instant of time with a 
very disordered landscape that is strongly detrimental to their mobility.
In this view, the presence of such unavoidable disordered molecular landscape constitutes the main and ultimate
limiting factor of the mobility in organic semiconductors.

The above observation calls for a change of paradigm from the conventional view of 
charge carriers being  weakly scattered 
by phonons, to one where charge carrier motion is hindered by a slowly varying, 
strongly disordered environment. 
\footnote{
Inter-molecular vibrations in organic semiconductors are only moderately coupled to the electron motion (inter-molecular 
electron-phonon coupling constants $E_P\lesssim 10 meV$ are roughly one order of magnitude lower than the 
couplings with intra-molecular modes), but their effect of on transport is 
strong owing to their large amplitude.  
This particular regime lies
beyond the limits of applicability of available electron-phonon coupling theories,
so that an alternative starting point (i.e. viewing them as a slow dynamical disorder) becomes more sensible.
In more rigorous terms, what is needed is a theory able to describe a 
weak or moderate coupling to pre-existent 
large molecular vibrations of thermal origin, to be contrasted with the well explored polaronic theories
which involve a strong coupling with carrier-induced molecular deformations.
}
Accordingly, one should observe even in the best organic semiconductors some of the characteristic features
of localization in disordered systems, and indeed there have been numerous reports of some form of 
localization of the carriers' wavefunctions in  organic semiconductors (see Sec. \ref{sec:exp} below).
Albeit slow, however, thermal molecular displacements are  by nature dynamic. 
This constitutes a fundamental difference
from the static chemical or structural disorder, which instead can cause 
full localization of the charges via a fully quantum process known as Anderson localization. \cite{LeeRMP1985}
For this reason, established theories of disordered conductors and semiconductors 
\cite{LeeRMP1985,BasslerReviewPSS1993,CoehoornPRB2005,FishchukPRB2009} 
cannot be directly applied to the problem, no more than the band, Marcus and polaron theories discussed above. 

The most striking consequence of dynamical disorder is that contrary to the static case, it
does not lead to the occurrence of thermally activated and exponentially suppressed mobilities, 
which are instead commonly observed in amorphous and disordered semiconductors. 
\cite{BasslerReviewPSS1993} Consequently, 
although a strong thermal molecular disorder is inherently present  in 
high quality organic crystals, 
the mobility in these systems exhibits a power-law temperature dependence 
which is  reminiscent of  semi-classical band-like behavior, 
in apparent contradiction with the disordered picture.
One of the main requirements for the theory is to solve this contradiction, accounting 
for both the intrinsic localization effects brought in by the strong molecular motions, and the 
apparent band-like  behavior of the mobility.

\paragraph{Transient localization: time for an overview. --}
The  ideas presented above constitute the basis of the  \textit{transient localization} scenario 
for  charge transport in organics. Starting from the pioneering works 
on a paradigmatic microscopic model 
that captures the essential aspects of the phenomenon,
\cite{DukeSchein80,Glarum63,FriedmanPR65,GosarChoiPR1966,MadhukarPostPRL1977}
a number of  recent theoretical works
have analyzed the problem by applying different numerical, 
analytical and phenomenological approaches. As a result of these recent studies, 
both  a solid theoretical framework and a more transparent physical picture of the charge transport 
mechanism are now emerging.
From the experimental point of view, several key experiments
are in agreement with the transient localization scenario, and an increasing number of results
is now discussed in terms of these ideas. 
\cite{Laarhoven-JCP08,SirringhausPSS2012,Chang-Troisi-SirringhausPRL2011,SirringhausNatMat2013,YadaAPL2014,MinderAdvMat2014}

Despite a number of review articles and books published on the subject of charge transport
in organic semiconductors, this important theoretical framework 
has not been comprehensively described in any review yet 
(although some aspects are mentioned in Refs. 
\cite{SirringhausPSS2012,TroisiPCCP08,TroisiChemSocRev2011}). 
Because it is now reaching its full theoretical maturity,
and in order to set the ground for more systematic experimental
confirmations, we consider that 
it is now timely to provide an overview of the transient localization scenario.

The present article has no ambition to be exhaustive on the different theories 
of charge transport in organic semiconductors. 
For these purposes, we refer the reader to existing reviews on the subject, e.g.
\cite{BredasChemRev04,DukeSchein80,SirringhausPSS2012,TroisiChemSocRev2011,CoropceanuChemRev2007,
Schweicher-IJCH2014,OrgElII,Shuai-review-AdvMat11,StafstromChemSocRev2010}.
Our aim here is  to focus on 
 the simplest and most studied model that captures the essential aspects of charge transport in  organic semiconductors: 
Eq. (\ref{eq:SSH}) --- which describes the motion of electrons on a one-dimensional
molecular stack, linearly coupled to inter-molecular vibrations. We shall
present the  different theoretical approaches that have been applied to it, and compare 
the different theories with existing experimental results.
For the sake of clarity, whenever possible we shall compare the results of different  approaches 
by keeping one given set of parameters corresponding to rubrene,
for which reliable theoretical estimates are available. 
This will allow us to benchmark the different methods
that have been applied to the calculation of the mobility, 
assessing how they perform on an experimentally relevant example.
\footnote{Due to the large amount of theoretical
results that have been produced in more than half a century of research,
it is not an easy task for the unexperienced reader to distinguish
which predictions for the mobility result from different 
microscopic models (say, coupling to intra-molecular vibrations vs. coupling to inter-molecular vibrations), 
which ones result from a given model but in 
different parameter regimes (e.g., coupling of electrons to molecular vibrations
in the strong coupling regime, $J\ll E_P$, or in the weak coupling regime, $E_P\ll J$), 
and which ones result when different approximate schemes are applied to the same 
model and with the same set of microscopic parameters.   
This article tries to answer the latter, focusing on  the specific case of Eq. (\ref{eq:SSH}).} 
For the same reasons exposed above, we choose rubrene as a  reference experimental material, 
which is one of the most studied high mobility organic semiconductors.


The outline of this article is the following. In Sec. \ref{sec:overview} we start
by briefly reviewing  a number of experimental results that are of direct relevance to
the present scenario. We then present the different possible  interactions at work
in  organic semiconductors and build the reference microscopic model that will be considered in the
following. In Sec. \ref{sec:early} we give a brief account on how traditional
theoretical approaches perform on such model. 
Sec. \ref{sec:transloc} describes  modern approaches that  
take quantum localization effects into account, leading to the 
transient localization scenario for charge transport. The relevance of transient localization to 
organic conductors and other degenerate systems is also discussed.
Sec. \ref{sec:out} provides a brief summary of the results presented in this overview.

\section{Experimental background and theoretical modeling}

\label{sec:overview}

\subsection{Experimental overview}
\label{sec:exp}

We present here some key experimental observations that are puzzling or contradictory from the point of view of
the available descriptions of organic semiconductors, 
calling for a new theory for charge transport.

\paragraph{Charge transport.--} 
The mobility values are low, falling below the Mott-Ioffe-Regel limit (see Sec. \ref{sec:breakdown}). 
Attempts to extract a mean-free-path from the band parameters yield values 
comparable to or shorter than the inter-molecular spacing. 
This is in contradiction with
the very assumptions of Bloch-Boltzmann band transport theory. Yet, 
in sufficiently pure samples and at sufficiently high temperatures
where extrinsic disorder effects are not crucial, the temperature dependence
of the mobility in the best  organic semiconductors appears to follow a power-law, as would be expected 
in conventional semiconductors. 

The Seebeck coefficient reported in Ref. \cite{PernstichNMat2008} also seems to be consistent with a "band-like" behavior.
The observation of a free-electron like Hall resistance \cite{PodzorovHallPRL2005} indicates that 
at least part of the
carriers' wavefunctions are extended over several molecules, as such delocalization is necessary for the Lorentz force
to have an effect on the electron motion. \cite{SirringhausPSS2012}

\paragraph{Photoemission.--}
Early experiments on the photoemission spectra of organic molecules in the gas phase  allowed to 
assess the presence of a sizable coupling between the molecular orbitals and the intra-molecular vibrations 
\cite{CoropceanuPRL2002},
confirming the values of the relaxation energy $E_P$ predicted by theoretical calculations. 
More recently, systematic angle-resolved photoemission experiments
(ARPES) performed in crystalline  organic semiconductors have shown beyond any doubt that such a coupling 
is not sufficient to destroy 
or even substantially shrink the electronic bands as would be predicted by 
polaron theory (cf. Sec. \ref{sec:hopping} below). 
Because well defined dispersive bands do exist in high-mobility  organic semiconductors, theories
which start from the molecular limit where the inter-molecular integrals $J$ are 
assumed to be small compared to the 
other energy scales should be taken with extreme care.

\paragraph{Electron diffraction.--}
The occurrence of large thermal molecular motions has been predicted to be a crucial factor in 
limiting charge transport in  organic semiconductors (cf. Sec. \ref{sec:transloc}).  
To address the question experimentally, the structural dynamics of crystalline samples of a pentacene derivative
have been investigated via electron diffraction \cite{SirringhausNatMat2013}.
Signatures of the thermal inter-molecular motion have been identified in the form of streaks in the diffraction patterns, 
which could be directly related to the presence of large inter-molecular sliding motions.
The average spread of dynamical  displacements has been estimated to be of the order of $\sim 0.1 \AA$ at 
room temperature.

\paragraph{Band tails.--}
A tangible consequence of such large molecular motions is that the electronic energy vs. momentum dispersion is
no longer sharply defined. Because the inter-molecular transfer integrals are themselves strongly 
fluctuating quantities, tails are expected to emerge in the density of states beyond the band edges.
Such tails can  be addressed experimentally from the analysis of the electrical characteristics of
field effect transistors. Recent works in this direction have shown that
even when extrinsic sources of disorder are removed, 
band tails of intrinsic origin remain, with an extension of few tens of $meV$. Such intrinsic tails
have been ascribed to the presence of thermal molecular motions.
\cite{KalbPRB2010,WillaJAP2013}

\paragraph{Optical conductivity.--}
Measuring the optical conductivity of charge carriers in organic semiconductors is experimentally challenging, as these  
must be injected either in a field-effect transistor (FET) geometry 
(in which case the absorption measurement is complicated by reflections at the interfaces)
or by optical pumping (which is in principle free from interface effects but drives the system out of equilibrium).
Despite the experimental difficulties, there have been several reports of the optical conductivity of excess carriers 
in rubrene.  Measurements by different groups \cite{FischerAPL2006,LiBasov07,YadaAPL2014}
agree in showing a markedly 
non-Drude behavior  at room temperature: the optical conductivity exhibits a finite frequency maximum in 
the far infra-red range,  
indicating a breakdown of semi-classical behavior,  which has been ascribed to some form of localization 
of the charge carriers. \cite{LiBasov07} We will come back to this crucial observation in Sec. \ref{sec:transloc}, when 
analyzing the predictions of the transient localization mechanism concerning the carriers'
optical response. 

\paragraph{Other probes of a finite localization length.--}
ESR measurements in different materials have pointed to the existence of a finite extension for the 
carrier wave-functions, typically on few molecular units. \cite{MarumotoPRL2006,MatsuiPRL2010,Marumoto11}
It is not clear however if the obtained lengths should be associated to  the presence of 
intrinsic thermal disorder or to  trapping by  extrinsic
sources of disorder. In any case, these findings 
provide strong indications that a true quantum localization process takes place in  organic semiconductors.
The optical charge modulated spectroscopy (CMS) measurements of Ref. 
\cite{SirringhausNatMat2010,Chang-Troisi-SirringhausPRL2011} 
in pentacene derivatives 
are also compatible with  the existence of a finite localization length,
coexisting with a band-like power-law dependence of the mobility.

\subsection{Microscopic mechanisms at work}
\label{sec:rev}

Several microscopic mechanisms have been considered in order to explain the 
low electronic mobilities in molecular organic semiconductors. These include
the coupling of electrons to low-frequency, inter-molecular vibrations; 
the coupling of electrons to high-frequency, intra-molecular vibrations;
the coupling to fast electronic polarization modes; 
the presence of bulk static disorder, of both chemical and structural origin as well as interface disorder
and polarization, when the material is placed in a FET geometry.
Although all these phenomena can certainly play a role, 
we shall focus in this article solely on the coupling to slow inter-molecular modes, 
which is now emerging as the main intrinsic mechanism limiting  charge transport in 
the best organic semiconductors.
Before moving on to the detailed model description of this phenomenon in Sec. \ref{sec:model}, 
we briefly comment here on the other mechanisms at work. 

\paragraph{High-frequency intra-molecular modes.--}
Intra-molecular modes mostly 
occur at high frequencies, $\hbar\Omega_0\sim 0.1-0.2 eV$, 
originating from the  stretching  of the strong covalent bonds inside the molecule.
The coupling to such  modes, if sufficiently strong, can lead to the formation of small polarons, 
\cite{Holstein-1959}
i.e. the self-trapping of  charge carriers by the molecular deformations that
they themselves create upon residing on a given molecule: a simple argument
predicts that when the energy gained
upon such a local deformation --- the relaxation energy $E_P$--- is larger than
the kinetic energy gained through delocalization on the periodic lattice, it
becomes advantageous for the electron to remain localized on an individual
molecule. Charge transport then occurs via hopping from site to site. Because 
at each hop the carriers must overcome an energy  barrier proportional to the
relaxation energy, the resulting mobility is exponentially activated with
temperature. 

There are two main reasons to exclude small polaron formation
in high-mobility  organic semiconductors. The most obvious is the experimental observation of "band-like"
mobilities, i.e. which decrease with temperature instead of increasing as in the 
thermally activated behavior expected for small polarons.
The second reason is that the calculated relaxation energies are insufficient to lead to small polaron formation 
in these materials: 
 it is known  theoretically that $E_P$ decreases with the size of the molecules
\cite{DevosPRB1998,CoropceanuPRL2002}, and the best  organic semiconductors are 
typically constituted of "large" molecules.
Furthermore, at the  values of $\Omega_0$ characterizing  organic semiconductors, the naive estimate $E_P\simeq J$ 
for polaron formation is too
optimistic, as phonon quantum fluctuations tend to delocalize the charges thus requiring
a higher value of $E_P$ in order to sustain a polaron. \cite{PiovraEPL1998}
The situation in the resulting intermediate regime of  relaxation energies  $E_P\sim J,\hbar \Omega_0$
cannot be described in simple terms, as it 
involves a subtle combination of coherent band transport
and incoherent hopping (see e.g. \cite{rhopolaron,Beljonne-SurfaceHopping-JPCLett2013,KataHolstein2015} 
for numerical results). Although in selected temperature intervals the  mobility 
around the polaron crossover could be compatible with the experiments,
it is unlikely that the "universal" power-law behavior observed in experiments 
can be explained assuming that all compounds lie in such fine-tuned crossover regime. 
In the regime of parameters which applies to high-mobility  organic semiconductors, 
high-frequency intra-molecular vibrations 
 weakly affect charge transport  via a modest renormalization
of the effective mass 
\cite{TroisiChemSocRev2011,VukmirovicPRL2012,orgarpes,exporgarpes}, but they 
are not the main limiting factor for the mobility.

\paragraph{Molecular polarization modes.--}
High-energy modes built from the electronic (excitonic) transitions in the molecules can also couple
to the carrier motion. While resulting in an appreciable
renormalization of the band parameters, \cite{GosarChoiPR1966,ZuppiroliEPL2004,PiconPRB2007} 
such  polarization modes of electronic origin are too fast to efficiently couple to
the carrier motion, and therefore do not directly affect the
carrier lifetime and scattering time. We shall assume that molecular polarization effects are already 
included in the definition of the inter-molecular transfer integrals of the model Eq. (\ref{eq:SSH}).

\paragraph{Static disorder.--}
Because our focus here 
is on the {\it intrinsic} mechanisms limiting the mobility in  organic semiconductors, 
we shall ignore the effects of structural  disorder \cite{BasslerReviewPSS1993,CoehoornPRB2005,FishchukPRB2009} or the influence of
chemical impurities and polymorphs.
As was studied  systematically both theoretically \cite{PRB12} 
and experimentally \cite{MinderAdvMat2014}, 
extrinsic sources of disorder can affect carrier transport 
even in the purest samples available nowadays, especially when these are placed in FET geometries. 
Static disorder causes the mobility to degrade at low temperature in the best  cases, 
and  wash out the
intrinsic transport regime completely in the worst cases, leading to a thermally activated behavior.  
We shall however consider here an ideal situation where all extrinsic sources of disorder have been removed.

\paragraph{Substrate polarization.--}
It has been shown that the presence of a polarizable 
substrate can strongly affect the transport properties in field-effect transistors. 
The carriers couple electrostatically 
to the polarization of the substrate, which provides an additional source of 
electron-phonon coupling, \cite{KirovaPRB2003,NatMat} 
leading to the possible formation of Fr\"ohlich polarons.
In the static limit, the polarization 
acts instead as an additional source of  disorder, due to the random arrangement of charged dipoles in the substrate. 
\cite{RichardsJCP2008,MinderAdvMat2014}
Both effects can convert the "band-like" temperature dependent mobility intrinsic to  organic semiconductors into
a thermally activated behavior, and will not be considered here.

\subsection{A paradigmatic model}

\label{sec:model}

\paragraph{Low-frequency inter-molecular modes.--}

\begin{figure}[th!]\center
\includegraphics[width=6.5cm]{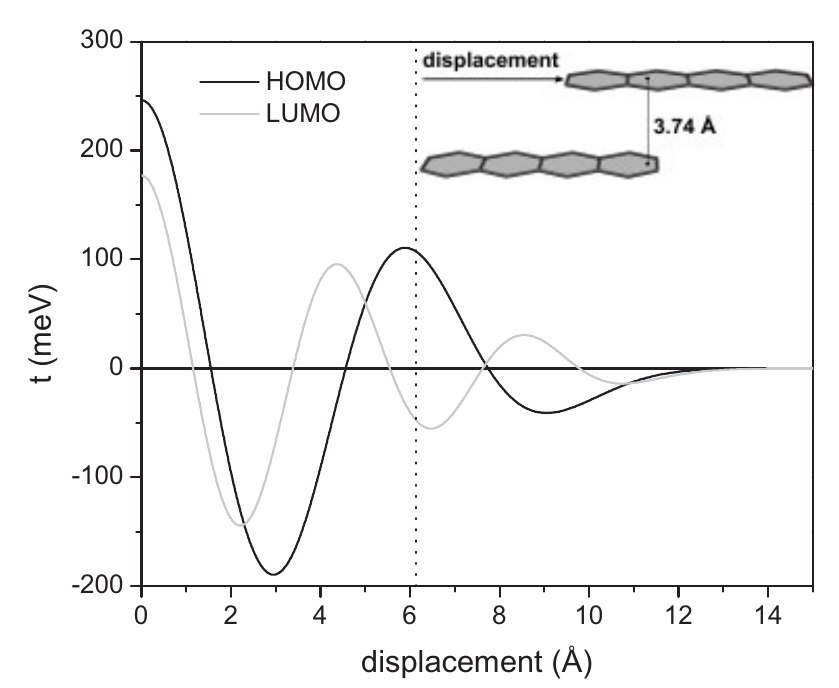}
\includegraphics[width=6.5cm]{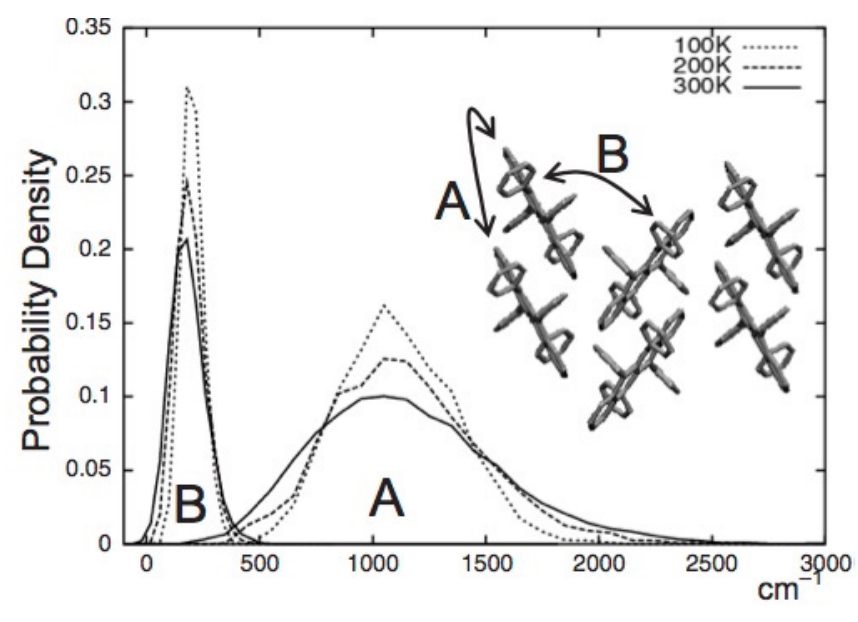}
\caption{Top: Evolution of the HOMO and LUMO transfer integrals as a
function of displacement, for a complex made of two tetracene backbones
stacked along the rubrene a direction. The dotted line indicates the equilibrium displacement in the rubrene crystal
(from Ref. \cite{DaSilvaAdvMat2005}). 
Bottom: the  distribution of transfer integrals between two different pairs of neighboring molecules in rubrene, 
calculated via molecular dynamics simulations, showing an essentially gaussian (thermal) probability
(from Ref. \cite{TroisiAdvMat2007}).}
\label{fig:Troisi}
\end{figure}

The coupling to inter-molecular vibrations  --- sometimes
called the Peierls- or Su-Schrieffer-Heeger-type interaction --- was considered already in the early days
of  organic semiconductors \cite{DukeSchein80,Glarum63,FriedmanPR65,GosarChoiPR1966,MadhukarPostPRL1977}, and revived later with 
the advent of organic field-effect transistors \cite{SinovaPRL2001}.
It is however only recently that it has become a cornerstone in the understanding charge transport in  organic semiconductors.
There exist now a number of {\it ab initio} calculations of the precise values of such coupling
by different
methods  and on different crystalline compounds  (see e.g. Refs.
\cite{GirlandoPRB10,CoropceanuChemRev2007,DaSilvaAdvMat2005,HannewaldBobbertAPL2004,HannewaldPRB2004,KatoJCP2002,
SanchezCarrera-JACS10}).
Without going through a review of all the extensive theoretical literature on the subject, 
we highlight in Fig. \ref{fig:Troisi} two observations which illustrate the essence 
of the microscopic mechanism and its implications.

Fig.  \ref{fig:Troisi} (top panel)  shows the evolution of the HOMO and LUMO transfer integrals between two adjacent 
organic molecules as a function of their relative displacement, calculated at the INDO level in a geometry which  
corresponds to the crystal structure of rubrene (from Ref. \cite{DaSilvaAdvMat2005}).
The figure illustrates two important facts. The first is that the transfer integrals between 
$\pi$ orbitals of neighboring molecules are typically on the order of $\sim 100 meV$ or below, leading to narrow electronic bands
whose width generally does not exceed $\sim 0.5 eV$. The second is that they 
are strongly varying functions of the displacement, exhibiting  
oscillations which (in the case illustrated here) 
are directly related to the periodic structure of the phenyl groups in the molecular backbone. 
The strong dependence of the transfer integrals on the displacement signifies that the electrons on molecular
orbitals are strongly coupled to the inter-molecular motions. 

Fig.  \ref{fig:Troisi} (bottom panel) shows the statistical 
distribution of transfer integrals between HOMO orbitals of rubrene, 
also computed at 
the INDO level, obtained upon averaging over time on molecular dynamics simulations
where the molecular positions in the crystalline matrix are allowed to thermally fluctuate 
(from Ref. \cite{TroisiAdvMat2007}). 
The two nonequivalent transfer integrals denoted A and B
exhibit essentially gaussian distributions of thermal origin, 
whose spread  increases with temperature, 
and becomes  comparable  to the mean value itself 
in the experimentally relevant temperature range.
Such large fluctuations, which are a key feature of organic solids, 
are a direct consequence of the weak van der Waals forces 
which bind the molecules together. As will be shown in the following paragraphs,
it is precisely the anomalously large magnitude of such fluctuations which is responsible for the 
original transport mechanism that characterizes organic materials.

\paragraph{Model Hamiltonian.--}

\begin{figure}[th!]\center
\includegraphics[width=7.5cm]{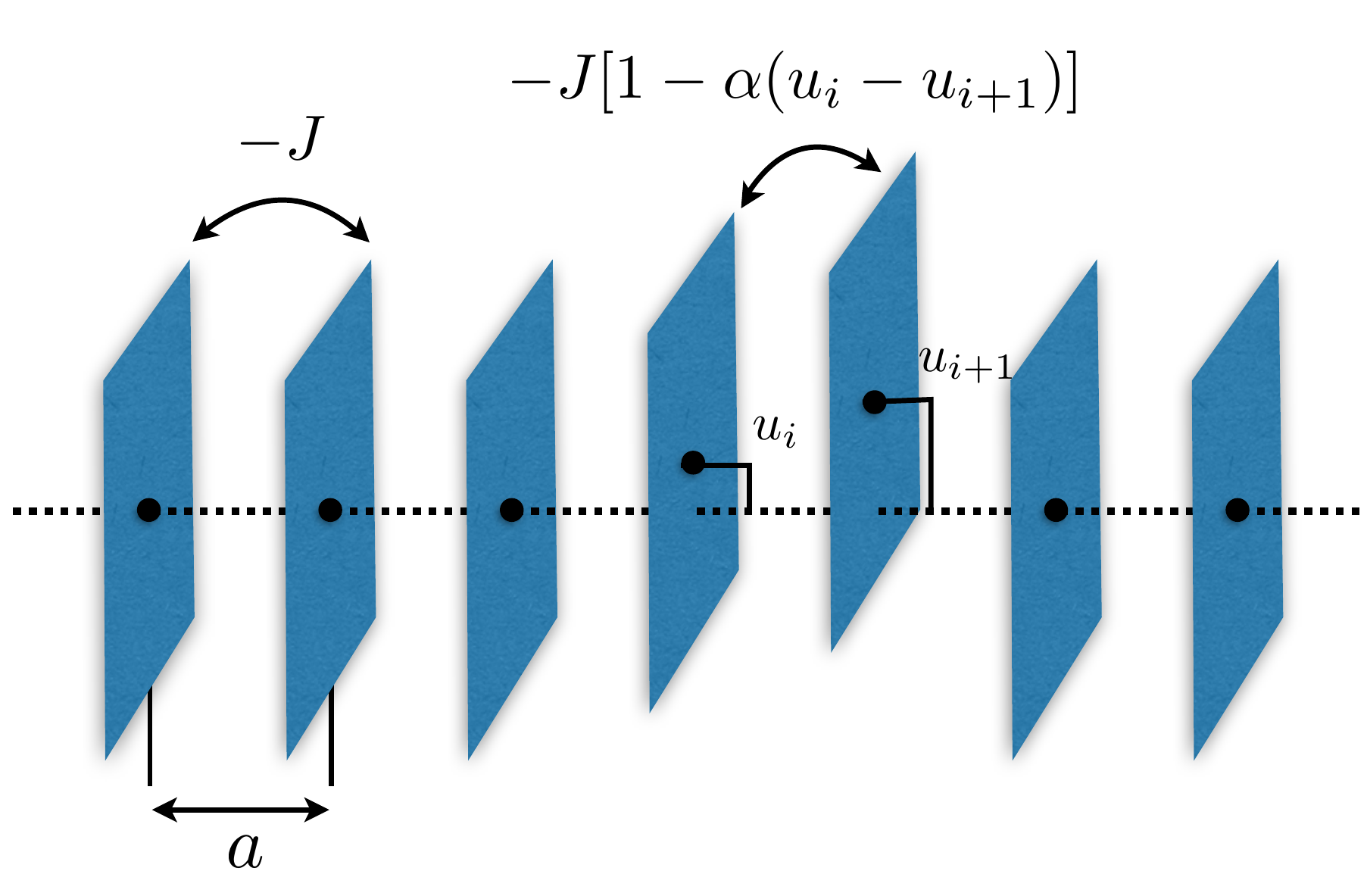}
\caption{Illustration of the model Eq. (\ref{eq:SSH}). The transfer integral $J$
between neighboring molecules is modulated by their relative displacement $u_i-u_{i+1}$  (readapted from 
\cite{TroisiPRL06}).}
\label{fig:model}
\end{figure}

The minimal model which accounts for the coupling to low-frequency molecular displacements 
in organic semiconductors
can be written in second quantization as  
\cite{TroisiPRL06,MadhukarPostPRL1977,reconcile09}
\begin{equation}
  H= - J\sum_i  [1-\alpha (u_i-u_{i+1})] \; (c^+_i  c_{i+1} + c^+_{i+1}  c_{i}) 
  + H_{vib},
  \label{eq:SSH}
\end{equation}
with 
\begin{equation}
   H_{vib}=\sum_i \frac{M \omega_0^2}{2} u_i^2+\sum_i \frac{p_i^2}{2M}.
  \label{eq:Hvib}
\end{equation}
The model is illustrated in Fig. \ref{fig:model}.

The first term describes electrons with creation (annihilation) operators
$c^+_i$ ($c_i$) hopping between molecular orbitals on neighboring sites along a one-dimensional chain (with lattice
parameter $a$), which are treated in a tight-binding approximation as appropriate to narrow-band solids. 
The key ingredient of the model is that 
the amplitude $J$ of inter-molecular transfer is modulated by the relative molecular displacements, $u_i-u_j$,
as $J_{ij}=J [1-\alpha (u_i-u_j)]$. The coordinates $u_i$ 
can represent in general either translational or rotational motions of the molecular units.
It is customary to take the dependence of the inter-molecular transfer integrals on the molecular coordinates
to  lowest (linear) order. This is justified in the relevant regime of thermal molecular motions (the average molecular 
displacements are estimated to be of the order of $0.1 \AA$ at room temperature, \cite{SirringhausNatMat2013} to be compared with the scale of Fig. \ref{fig:Troisi} (top)).
As a consequence, deviations from linearity
are weak, and are at the origin of the small deviations from the purely gaussian shape in  the statistical distributions of 
 Fig. \ref{fig:Troisi} (bottom).

The second term, 
$H_{vib}$,
describes the harmonic vibrations of the
molecules around their equilibrium positions. 
The characteristic frequencies of the inter-molecular vibrations are in the range  $\hbar \omega_0 \lesssim 10 meV$.
Such low values, which result from both the weak inter-molecular forces and the large weights of the moving molecular units,
allow for a classical treatment of the molecular motions as long as  $\hbar \omega_0 \lesssim k_BT$ (see below).

To the best of our knowledge, the above model was first introduced in the early days of organic 
semiconductor research
in the k-space form Eq. (\ref{eq:SSHk}) given below. \cite{Glarum63,FriedmanPR65}
The interaction term is formally analogous to the popular Su-Schrieffer-Heeger (SSH) model 
\cite{SSH+KivelsonRMP1088} that describes 
the properties of conjugate polymer chains, although it is applied here to a 
different physical situation.
The first important difference is that in the  SSH model, 
acoustic phonons are considered instead of the
dispersionless vibrations of Eq. (\ref{eq:SSHk}). The qualitative features of the model 
in the case of acoustic phonons are different, see e.g. Refs.
\cite{CSGSSHPRB1997,ZoliSSHacuPertPRB2002}.
Secondly, here a low density of injected carriers is considered
while in polymers the electron density  is large and the electron system is degenerate 
(the bands are half-filled by construction,  and the SSH interaction
is itself responsible for dimerization of the structure and for the consequent opening of a gap at the Fermi energy).
The final difference is quantitative, as both the inter-molecular transfer integrals $J$ and the 
scale of the inter-molecular vibration frequencies  in small-molecule organic semiconductors are more than one order 
of magnitude smaller than in polymers. 


The electronic properties of the model Eq. (\ref{eq:SSH})
can be expressed in terms of the temperature $T$ and two dimensionless coupling parameters: 
the electron-lattice coupling strength 
$\lambda=\alpha^2 (\hbar/2M\omega_0) (J/\hbar\omega_0)$ and the adiabatic ratio $\hbar\omega_0/J$.
Unless otherwise specified, we shall consider in what follows
the values of microscopic parameters evaluated in Ref. \cite{TroisiAdvMat2007} for the direction 
of highest conduction in 
rubrene, i.e. $J=143meV$, $\omega_0=6.2meV$ ($\hbar\omega_0/J=0.044$)
and $\lambda=0.17$, which fall in the typical range of parameters of high mobility organic semiconductors


\section{Semi-classical and hopping theories}
\label{sec:early}

\begin{figure}[th!]\center
\includegraphics[width=8cm]{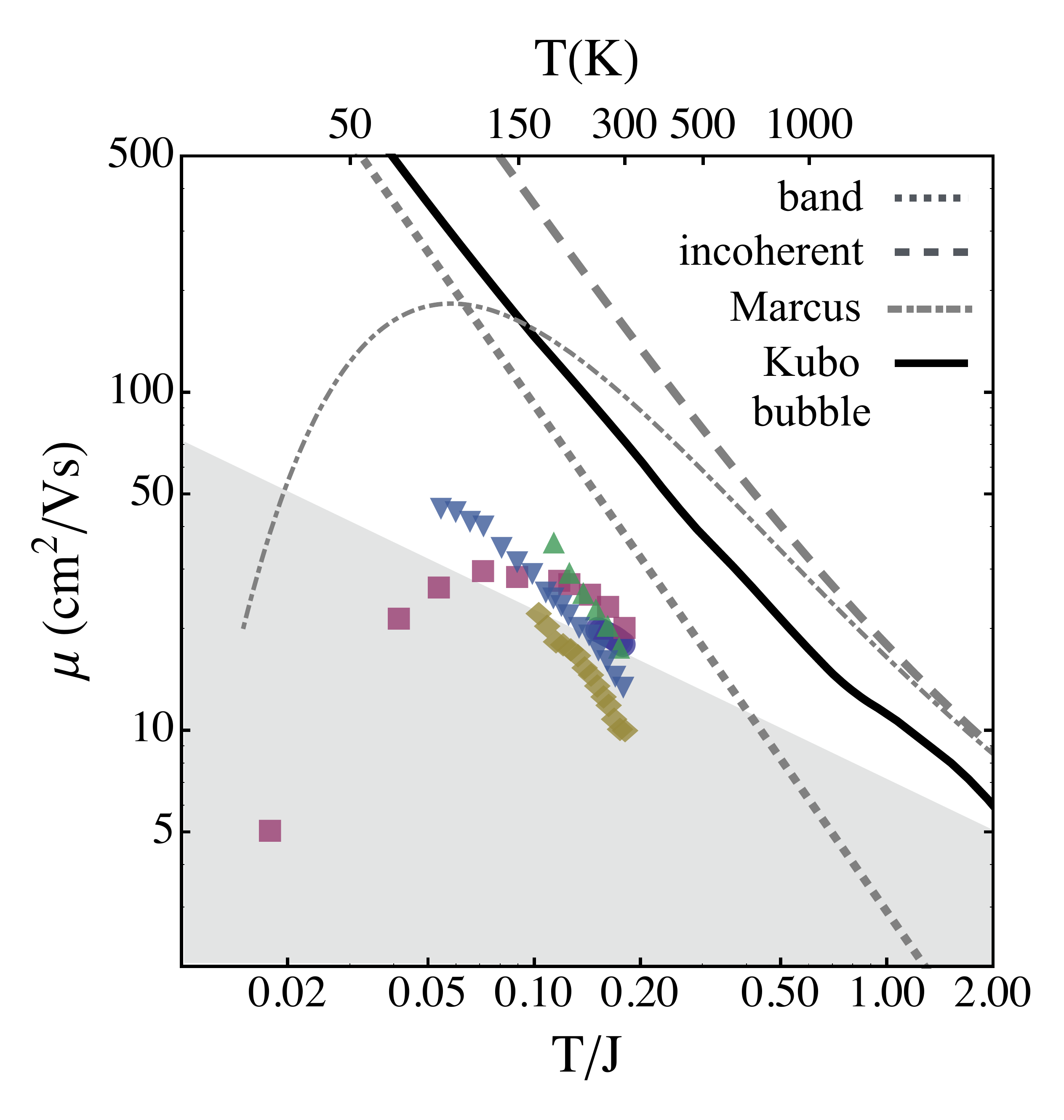}
\caption{Mobility calculated from Eq. (\ref{eq:SSH}) using semi-classical and hopping theories: band theory, 
Eq.(\ref{eq:loT});
incoherent limit, Eq. (\ref{eq:musat}); Marcus theory, Eq. (\ref{eq:Marcus}) and Kubo formula in the 
bubble approximation, Eq. (\ref{eq:mobzero}). The microscopic parameters are those 
given at the end of Sec. \ref{sec:model}.
The gray shaded area shows the Mott-Ioffe-Regel limit defined by Eq. (\ref{eq:muMIR}).
The data points represent different experimental measurements on rubrene FETs: disks \cite{SakaiPRB12}, squares \cite{PodzorovPRL2004}, diamonds \cite{PodzorovHallPRL2005}, 
up triangles \cite{NatMat}, down triangles \cite{FrisbieJPC2013}.   
The theoretical curves all overestimate the experimental mobilities,
suggesting that some important microscopic phenomenon is missing in the description.}
\label{fig:semic}
\end{figure}

This Section presents a brief description of how standard approaches perform 
on the model Eq. (\ref{eq:SSH}). 
The  results for the temperature dependence of the mobility, when available either in analytical or numerical 
form,  are summarized in Fig. \ref{fig:semic} and compared with existing measurements in rubrene FETs. 
Fig. \ref{fig:semic} shows that  "standard" methods generally overestimate the mobility: 
the  values calculated at room temperature fall in the range $\mu=50-200 cm^2/Vs$, 
while the measured  mobilities in rubrene  are around $\mu=10-20 cm^2/Vs$.
Such disagreement tells us that  
a fundamental mechanism of reduction of the mobility is missing in these theories.
While one could try to bring the theoretical values closer to the 
experimental range by including a number of other interactions among the ones enumerated in Sec. \ref{sec:rev}, 
we find it unlikely that these effects will restore 
a proper agreement if they are treated correctly. 
As will become clear in Section \ref{sec:transloc}, the failure of standard theories
is due to the fact that they  
do not take quantum localization effects  into account. 
Including these effects not only restores  the quantitative agreement with the experimental mobilities, but also 
solves the experimental puzzles identified in the preceding Section concerning the duality between 
the extended and localized character of the charge carriers.

\subsection{Semi-classical approaches}
\label{sec:semic}
\subsubsection{Band transport}

Bloch-Boltzmann transport theory starts from the solution of the electronic problem in 
an unperturbed, perfect lattice. 
In this limit the electrons form Bloch waves
identified by a well-defined momentum $k$ and energy $\epsilon_k$.
Scattering to impurities or lattice vibrations
 is included as a perturbation, 
by introducing a transport relaxation time $\tau^{tr}_k$ for the Bloch eigenstates. 

To work in momentum space, 
one proceeds by rewriting  Eq. (\ref{eq:SSH}) as
\begin{eqnarray}
  H&=& \sum_k \epsilon_k c^+_k  c_k   -\frac{1}{\sqrt{N}}\sum_{kq} g_{k,k+q} c^+_{k+q}  
  c_k (a^+_q+a_{-q}) \nonumber \\
   & & + \sum_q \hbar \omega_0   (a^+_qa_q+1/2)
  \label{eq:SSHk}
\end{eqnarray}
with  $\epsilon_k=-2J\cos k a$  the eigenenergies of the unperturbed one-dimensional chain, 
$g_{k,k+q}=2ig[\sin(k+q)a-\sin ka]$  the Fourier transform of the 
electron-vibration interaction, with $g=(\alpha J)\sqrt{\hbar/(2M \omega_0)}$, and
$a^+_q,a_q$ the phonon creation and annihilation operators. 
Note that with these definitions, the  
dimensionless coupling constant takes the  form $\lambda=g^2/(\hbar \omega_0J)$.

In the Boltzmann description of electronic transport in nondegenerate
semiconductors,  the mobility is  expressed as \cite{Mahan} 
\begin{equation}
  \label{eq:Boltz}
  \mu(T) = \frac{e}{nk_BT} \sum_k  v_k^2 \tau^{tr}_{k}
  e^{-(\epsilon_k-\tilde \mu)/k_BT} ,
\end{equation}
where $\tau^{tr}_{k}$ and $v_k$ are respectively the transport
scattering time and the band velocity for electrons of momentum $k$, 
$\tilde \mu$ is the chemical
potential and $n=\sum_k  e^{-(\epsilon_k-\tilde\mu)/k_B T}$ the thermal
population of carriers. In the quasi-elastic limit where the phonon
frequency sets the smallest energy scale in the problem,
$\hbar\omega_0\ll k_BT,J$ the scattering time is defined as
\begin{equation}
  \label{eq:trscat}
  1/\tau^{tr}_{k}= \frac{2k_BT}{\hbar \omega_0}  \int dq \; g^2_{k,k+q}
(1-\cos \theta_{k,k+q}) \; \delta(\epsilon_k-\epsilon_{k+q})
\end{equation}
with  $\theta_{k,k+q}$ the angle between the
incoming and outcoming momentum states. 
Reminding that the coupling matrix element $g_{k,k+q}\propto g$ and using the definitions given after Eq. (\ref{eq:SSHk}), 
we see that the scattering rate $ \hbar /\tau^{tr}_{k} \propto \lambda k_BT$ (cf. Ref. \cite{reconcile09}).
The present theory is consistent provided that the scattering rate 
is small compared to the bandwidth, i.e. $\lambda k_BT \ll J$.

The integral in Eq. (\ref{eq:trscat}) 
can be carried out analytically for the present model, yielding
\begin{equation}
  \label{eq:Boltz-clas}
  \mu_{\textrm{band}}=\frac{\mu_0}{16\pi
    \lambda}\frac{J}{k_BT}\frac{\sinh(2J/k_BT)}{I_0(2J/k_BT)}, 
\end{equation}
with  $I_0$ the modified Bessel function.
In the above expression, the absolute units of mobility 
are set by the prefactor $\mu_0=ea^2/\hbar$, which equals $7cm^2/Vs$ in the direction of highest mobility of rubrene.
Note that the above result can be straightforwardly
generalized to the case of a finite vibrational frequency $\omega_0\neq 0$, in which case the mobility is seen to
rise exponentially at temperatures $k_BT\ll \hbar\omega_0$ due to the
dispersionless (gapped) nature of molecular vibrations. 
%

In the limiting cases of temperatures much lower or much higher than
the bandwidth, the above
expression reduces to the following power laws: \cite{Glarum63,FriedmanPR65}
\begin{eqnarray}
  \label{eq:loT}
  \mu_{\textrm{band}}&=&\frac{\mu_0}{8 \sqrt{\pi} \lambda} \left(
    \frac{J}{k_BT}\right)^{3/2} \;\;\;\; k_BT\ll 2J \\
  \mu_{\textrm{band}}&=&\frac{\mu_0}{4\pi \lambda} \left( \frac{J}{k_BT}\right)^2
  \;\;\;\; k_BT\gg 2J.
 \label{eq:hiT}
\end{eqnarray}
The low temperature result is illustrated as the dotted line in Fig. \ref{fig:semic}. 
The high temperature result is not shown, as it applies outside the relevant range in experimental systems: 
taking the representative value $J=143meV=1660K$ for rubrene from  Sec. \ref{sec:model}, 
the high temperature limit would apply at $T\gg 2J \sim 3000K$.


\subsubsection{Mott-Ioffe-Regel condition}
\label{sec:breakdown}

In the large scattering regime, the theory of band conduction
presented above breaks down due to the progressive loss of momentum
conservation. In non-degenerate semiconductors, this happens when the apparent mean-free-path $\ell$ for band electrons
reduces to values comparable or below the inter-molecular spacing $a$. 
The Mott-Ioffe-Regel (MIR) condition $\ell\simeq a$ thus provides a lower bound for the applicability 
of Bloch-Boltzmann theory. 

To determine the  value of the mobility corresponding to the MIR condition,
we start from the semi-classical Drude expression 
\begin{equation}
  \label{eq:Drude}
  \mu=\frac{e\tau}{m^*}.
\end{equation}
 Introducing the mobility units $\mu_0=ea^2/\hbar$ and 
the band mass $m^*=\hbar^2/(2Ja^2)$ for the considered one-dimensional lattice, 
we  rewrite Eq. (\ref{eq:Drude}) as
\begin{equation}
  \label{eq:Drudemu0}
  \mu=\mu_0 \frac{2J}{(\hbar/\tau)}.
\end{equation}
We now recall that the mean-free-path is the length over which the carriers
diffuse between successive scattering events separated by $\tau$, which  can be written as 
$\ell\sim \sqrt{D\tau}$, with $D$ the diffusion constant. Using the semi-classical expression
$D=\langle v^2 \rangle \tau$  and the equipartition principle $\langle v^2 \rangle = k_BT/m^*$, 
and imposing the condition
$\ell/a=1$ we arrive at 
\begin{equation}
  \label{eq:muMIR}
  \mu_{MIR}\sim \mu_0 \left(\frac{2J}{k_BT}\right)^{1/2},
\end{equation}
which is illustrated in Figs. \ref{fig:semic} and \ref{fig:mobility}.
Even in a "high mobility" organic semiconductors such as rubrene, the condition $\mu>\mu_{MIR}$ is experimentally violated 
at room temperature.
Taking  $J=143meV$ and $T=300K$ yields a MIR value  
$\mu_{MIR}\simeq 23cm^2/Vs \simeq 2\mu_0$. This corresponds to the typical 
experimental values in rubrene clean samples, which are precisely in the range $10-20cm^2/Vs$.
A  similar estimate was obtained
in Ref. \cite{ChengJCP2003} based on the actual band structure of oligoacene crystals 
(see also \cite{VukmirovicPRL2012,PRB14}
for a calculation of the scattering time).
Comparing this estimate with Eq. (\ref{eq:Drudemu0}) above we see that 
the MIR condition  roughly  coincides with the condition  that the elastic scattering time  
becomes shorter than the intermolecular transfer time, $\tau \lesssim  \hbar /J$.  \cite{PRB14}

We also mention here that the breakdown of Bloch-Boltzmann theory and the resulting resistivity saturation 
has  been thoroughly explored in the case of degenerate systems, leading to the concept of
"bad metals" (cf. Refs. \cite{MillisPRL1999,GunnarssonNat2000}). 
In that case the MIR condition becomes $\ell\simeq 1/k_F$ instead of $\ell\simeq a$, with $k_F$ the Fermi wavevector.

\subsubsection{Fully incoherent limit}
\label{sec:incoherent}


The semi-classical transport theory was generalized to the large scattering case by Gosar and Choi 
\cite{GosarChoiPR1966} and later by Sumi, 
\cite{SumiSSHJCP79} based on the Kubo formula for the electrical conductivity. \cite{Kubo57,Mahan} These authors 
observed that, even remaining in the spirit of semi-classical transport which neglects quantum localization effects altogether
(see below and Sec. \ref{sec:transloc}),
a different regime  of charge transport sets in when  the fluctuations of the
transfer integrals exceed their mean value $J$, where the bands themselves 
are no longer well-defined. 

The  transfer integrals fluctuate as a direct consequence of the 
thermal molecular motions, as encoded in  Eq. (\ref{eq:SSH}).
To estimate their spread we
observe that in the harmonic approximation
each vibrational mode $u_i$ is gaussianly distributed with fluctuations proportional to $k_BT$. 
Treating $u_i$ and $u_j$ as independent variables, and collecting the numerical factors, 
yields for the inter-molecular transfer integrals 
 $J_{ij}$ a gaussian distribution whose variance is $s^2=4 \lambda J k_BT$.  
The condition for fully incoherent transport is therefore
$s\gtrsim J$, i.e. $k_BT\gtrsim J/(4\lambda)$, which is
opposite to the band-transport condition given before Eq. (\ref{eq:Boltz-clas}).

Although it is true that thermal molecular fluctuations are large in organic semiconductors,
this extreme limit is never reached experimentally. For example,  
taking $J=1660K$ and $\lambda=0.17$ for rubrene from  Sec. \ref{sec:model} implies
that the fully incoherent limit would be reached for $T\gtrsim 2500 K$.
It is however important to describe this mechanism here because,
as will be shown via the comprehensive treatment of 
Sec. \ref{sec:reconcile},
both the diffusion of incoherent states and the coherent transport of Eq. (\ref{eq:loT})
are predicted to participate to semi-classical charge transport  at 
the experimentally relevant temperatures.

In the limit $s\gg J$, the weakly scattered 
Bloch waves with a well-defined momentum $k$ that enter into Eq. (\ref{eq:Boltz}) 
are replaced by a fully incoherent density matrix describing a
diffusion from site to site as in a classical random walk. 
In mathematical terms, the fact that all informations on the energy dispersion of momentum states are lost translates
into  the fact that the spectral function   $\rho(k,\omega)$ describing the quasiparticle excitations 
becomes $k$-independent, i.e.  $\rho(k,\omega)\to \rho(\omega)$. 
The spectral function becomes a local quantity,  signalling the loss of  coherence between different
molecular sites. It can be easily shown that it tends to a gaussian of variance $s^2=4 \lambda J k_BT$ itself
\cite{GosarChoiPR1966,SumiSSHJCP79} 
\begin{equation}
  \label{eq:gaux}
  \rho(\omega)=\frac{1}{\sqrt{2\pi s^2}}e^{-\omega^2/2s^2}.
\end{equation}
This spectral function can be directly inserted in the Kubo formula \cite{Kubo57,Mahan}, 
in the simplest approximation where the current-current response function is taken to be 
a convolution between two single-particle propagators --- the so-called  "bubble" form \cite{rhopolaron}. This 
yields the following mobility:
\begin{equation}
  \label{eq:Kubo}
  \mu_{\textrm{inc}}=\mu_0 \frac{\pi \langle J_{ij}^2\rangle}{x k_BT} 
  \int d\omega  \rho(\omega)^2  e^{-(\omega-\tilde{\mu})/k_BT}
\end{equation}
with $x=\int d\omega  \rho(\omega)  e^{-(\omega-\tilde{\mu})/k_BT}$ the thermal population of non-degenerate carriers, 
$\tilde{\mu}$ being the chemical potential.
The prefactor $\langle J_{ij}^2\rangle=J^2(1+4\lambda k_BT/J)$  accounts for
the temperature dependence of the mean square  of the inter-molecular transfer integrals. 
Its increase with respect to the 
reference value $J^2$ represents the thermally assisted tunneling processes caused by the fluctuations of the molecular positions, 
which set in and favor 
incoherent hopping at temperatures $k_BT\gtrsim J/4$, i.e. essentially where the present incoherent limit applies.
The importance of such vibrationally-assisted hopping was first recognized 
by Gosar and Choi in Ref. \cite{GosarChoiPR1966}.

Performing the integral above leads to 
\begin{equation}
  \label{eq:musat}    
\mu_{\textrm{inc}}=\mu_0 \sqrt{\frac{\pi}{8 \lambda}} \left(
     \frac{J}{T}\right)^{3/2}   \left( 1+4\lambda \frac{k_BT}{J}\right). 
 \end{equation}
Similarly to the band conduction of Eqs. (\ref{eq:loT}) and (\ref{eq:hiT}),
the fully incoherent diffusion mechanism of Gosar and Choi also results in a metallic-like, power-law 
temperature dependence of the mobility, which is displayed in Fig. \ref{fig:mobility} as a dashed line.
In the high temperature regime where the fully incoherent limit applies,  
the power law  $\sim T^{-3/2}$ that comes from the progressive energy 
spread of electronic states  (the first term in Eq. (\ref{eq:musat})) is converted into 
$\sim T^{-1/2}$ due to phonon-assisted tunneling. 

We stress here that although Eq. (\ref{eq:musat}) describes a regime where band transport is completely destroyed, 
it remains essentially  a semi-classical approximation to charge transport. 
The semi-classical nature of the theory
is implicit in having expressed the Kubo formula  in Eq. (\ref{eq:Kubo})
as a convolution integral, therefore neglecting 
all particle-hole interferences (the so-called vertex corrections) which  
are responsible for quantum localization processes. \cite{LeeRMP1985}

\subsection{Hopping and polaron-based theories}
\label{sec:hopping}

The failure of semi-classical approaches and the evidence of 
short mean-free-paths has been traditionally considered as an indication 
of polaronic localization of the carriers, which has led
to the widespread application of  hopping transport theories to crystalline organic semiconductors.
Such theories have been used despite the fact that,  
as was noted in Sec. \ref{sec:rev}, for  polaronic localization to occur
the  polaron energy should be the largest scale in the problem, which is not the case 
for the most conductive organic crystals such as pentacene or rubrene. 
\cite{VukmirovicPRL2012,TroisiJCP2011} 
We review here the predictions of these approaches when applied to the model Eq. (\ref{eq:SSH}).

\subsubsection{Marcus theory}
Marcus theory describes the hopping motion of charge carriers which are self-trapped by their induced
molecular deformations.
The corresponding mobility is calculated starting from the  transfer rate between two adjacent molecules, 
assuming that 
the quantum coherence is lost after each hopping event,
and takes the following thermally activated form:
\begin{equation}
\label{eq:Marcus}
\mu(T)=\mu_0 \ \langle J^2_{ij} \rangle 
\left(\frac{\pi}{4\varepsilon_{r} (k_BT)^{3}}\right)^{1/2} e^{-\varepsilon_{r}/4k_BT}.
\end{equation}
The quantity $\varepsilon_{r}=2\lambda J$ is the inter-molecular reorganization energy, 
which determines the energy barrier, and is proportional to the  coupling constant $\lambda$ 
($\varepsilon_P=\varepsilon_{r}/2=\lambda J$ is the corresponding relaxation energy associated
to inter-molecular vibrations). 
As in Eq. (\ref{eq:musat}), $\langle J^2_{ij} \rangle$ is the mean square   of the 
inter-molecular transfer integrals, and we have specialized to one space dimension. 
This behavior is illustrated in Fig. \ref{fig:semic} as a dash-dotted curve.


Marcus theory relies on the assumption that the molecular energy scales 
(in the present case the 
inter-molecular relaxation energy $\varepsilon_P$ and vibrational energy $\hbar \omega_0$) are much 
larger than the
transfer integral $J$. Even when these 
conditions are not fulfilled, however, Eq. (\ref{eq:Marcus}) 
exactly recovers the incoherent conductivity Eq. (\ref{eq:musat}), provided that the temperature 
is much higher than the activation barrier, $k_BT\gg \varepsilon_r/4$ (cf.  Fig. \ref{fig:semic}, where $\varepsilon_r/4= 0.085J$).
This agreement with semi-classical transport theory at high temperatures provides it with a certain interpolating
power and somehow justifies its use when addressing qualitative trends between different compounds.
\cite{Schweicher-IJCH2014}


We note that Eq. (\ref{eq:Marcus}) is  formally equivalent to the mobility obtained in  small polaron 
theory within the 
so-called  Holstein molecular crystal model \cite{Holstein-1959} in
 the non-adiabatic limit $J\ll\hbar
\omega_0$ and at sufficiently high temperatures, $k_B T \gg J$.  For an overview of the different transport
regimes of small polarons we refer the reader to  Refs. \cite{rhopolaron,KataHolstein2015,AustinMott}.

\subsubsection{Small polaron theory and extensions}


Early theories of small polaron transport based on the local 
Holstein-type coupling \cite{Holstein-1959} have  been adapted to non-local electron-phonon coupling
typical of organic systems 
\cite{HannewaldPRB2004,MunnSilbeyJCP1985,KenkrePRL89}. 
In a series of works \cite{HannewaldBobbertAPL2004,HannewaldPRB2004,KenkrePRL89} a power law
behavior $1/T$ of the mobility has been predicted to occur under the conditions that the electron-phonon
coupling is not too large and that  the activated regime is not reached in the 
temperature range of interest.
In subsequent  works \cite{OrtmannAPL2008,OrtmannPRB2009} the theory has
been extended to recover the weak coupling
case as a limiting behavior. A careful analysis of the terms arising from the structure of the
current-current correlation function allows to separate a coherent and an
incoherent contribution. The sum of these two contributions can yield 
a power-law different from $1/T$.

To critically examine these theories and their possible application to organics
we need to go further into the technical details of the calculations. To this aim we 
follow the treatment of Ref. \cite{HannewaldPRB2004} and adapt it 
explicitly to the model
Hamiltonian Eq. (\ref{eq:SSH}) by formulating it in real space. 
All the above theories are ultimately based on the so called
polaron \cite{Holstein-1959} or "Lang-Firsov" transformation, \cite{LangFirsovZETP1962,LangFirsovJETP1963} generalized to 
$q$-dependend electron-phonon matrix elements as in Eq. (\ref{eq:SSHk}). 
Such transformation is devised to formally cancel the electron-vibration interaction term in the Hamiltonian,
which is achieved by changing the electron operators into "dressed" polaron operators and shifting the phonon operators
by a constant term.
This is done in practice by introducing the unitary
trasformation $U = e^{S}$ where 
\begin{eqnarray} S&=&-\sum_{<i,j>} s_{i,j}c^\dagger_i c_j \label{eq:LFPeriels}  \\ 
s_{i,j}&=& \sum_q g^{i,j}_q (a_q-a^{\dagger}_{-q})\nonumber \\ 
g^{i,j}_q&=&\frac{1}{N\omega_0}\sum_k g_{k,k+q}e^{ik(R_j-R_i)}e^{-iqR_i} .
\nonumber
\end{eqnarray} 
The transformed Hamiltonian can be rewritten in terms of the original operators as
\begin{equation}
U H U^{\dagger} = -\sum_{i,j} {\bar J}_{i,j} c^{\dagger}_i c_j +\omega_o\sum_q a^\dagger_q a_q
+ H_{res}
\label{eq:polaronSSH} 
\end{equation}
where ${\bar J}_{i,j}=J\sum_{k,l} [e^{-s}]_{i,k} c^\dagger_k c_l
[e^{s}]_{l,j}+\delta_{i,j} \varepsilon_P$ contains a renormalized kinetic energy operator
and  a polaronic shift in the energy 
(s being the matrix whose elements $s_{i,j}$ are defined in Eq. (\ref{eq:LFPeriels})).

Contrary to the original Lang-Firsov transformation applied to the case of  local Holstein-type coupling,  Eqs.
(\ref{eq:LFPeriels}) are non-local, and 
do not  entirely remove the electron-phonon interaction because
they do not yield a simple
shift of the phonon operators.  Indeed, we have
$U a_q U^\dagger = a_q+\sum_{i,j} [s,a_q]_{i,j} c^{\dagger}_i c_j + \Delta a_q\nonumber$ 
where the shift is $[s,a_q]_{i,j}=\frac{g^{i,j}_{-q}}{\omega_0}$ and $\Delta
a_q$ contains non-linear terms in the phonon operators $a_q$ and $a^\dagger_q$
(a similar equation holds for $a^+_q$).
These are responsible for the residual
term $H_{res}$ in Eq. (\ref{eq:polaronSSH}).
 This term has been argued to be irrelevant
provided that the electron-phonon coupling is not too large \cite{HannewaldStojanovicJPC2004} and
it is usually neglected in the polaronic treatment of the mobility.
\cite{HannewaldPRB2004}

The kinetic energy terms in Eq. (\ref{eq:polaronSSH}) contain phonon operators to
all orders in $g_{k,k+q}$, which cannot be treated exactly 
(this is where the unitary transformation has moved the original interaction terms). 
The next step in the calculation is thus to implement an
approximation devised by Holstein \cite{Holstein-1959} which treats the polaron kinetic energy
by averaging the phonons over their thermal state. This leads to the following renormalized band dispersion:
\begin{equation}
{\bar \epsilon}_k = \frac{1}{\sqrt{N}}\sum_{i,j} e^{-ik(R_i-R_j)} \sum_{<n,m>}\langle [e^s]_{i,n} [e^{-s}]_{m,j} \rangle.
\label{eq:bandnarrowing}
\end{equation}
A further common approximation is to take only the nearest neighbor terms in the sum of Eq.
(\ref{eq:bandnarrowing}).
The  renormalized band dispersion resulting from Eq.
(\ref{eq:bandnarrowing}) has been evaluated using 
realistic band structure parameters obtained from density functional theory (DFT) in Ref. \cite{HannewaldPRB2004}.
The main qualitative results are basically  the same as obtained in the  model Hamiltonian Eq.
(\ref{eq:SSH}), and which are well known since the works of Holstein: 
above a crossover temperature $k_B T_c \simeq \hbar \omega_0$,
the coherent polaron bandwidth decreases with temperature  
due to the exponential factors in Eq. (\ref{eq:bandnarrowing}).

The calculation of the mobility now proceeds through the Kubo formula \cite{Mahan,Kubo57} 
(see e.g. Refs. \cite{Holstein-1959,OrtmannPRB2009}).
Due to the decoupling between polaron and phonons achieved by Eq. (\ref{eq:polaronSSH}),
the current-current correlation function is factored   into  products of
bosonic and fermonic correlators. We can write schematically
$\langle \hat{J}(t)\hat{J}(0)\rangle=\langle \hat B(t)\hat B(0)\rangle\langle \hat{J}^{(p)}(t)\hat{J}^{(p)}(0)\rangle$,
where the $\hat B(t)$ contain only phonon operators and
$\hat J^{(p)}(t)$ is the polaron current operator. 
The same structure applies to the single particle propagator, as shown in Ref. \cite{RanningerPRB1993}.
Writing $\langle \hat B(t) \hat B(0)\rangle=\langle \hat B(t)\hat B(0)\rangle-\langle \hat B^2(0)\rangle+\langle \hat B^2(0)\rangle$ 
it is possible to recast 
the current-current correlator
as  as a sum of two contributions: (i) a coherent part which is
that of unscattered free polarons, 
$\langle \hat B^2(0)\rangle\langle \hat J^{(p)}(t)\hat J^{(p)}(0)\rangle$, 
and which incorporates 
the terms responsible for the band-narrowing;
(ii) a remainder interpreted as an incoherent part which includes polaron-phonon scattering.
At this level of approximation, i.e.  as long as  $H_{res}$ is neglected, 
the polaron-phonon scattering comes 
from the time dependence of the bosonic  operators
$\hat B(t)$, 
which however does not enter in the coherent part. As a consequence, 
the latter is formally divergent and an \textit{ad hoc} regularization 
parameter must be introduced in the theory. \cite{Holstein-1959,OrtmannPRB2009}
When this is done, a 
 crossover between the two contributions (i) and (ii) occurs  as
the temperature increases above $k_B T \simeq \hbar \omega_0$).
When calculated for the typical intra-molecular modes in organic semiconductors, this locates the
crossover around $50-100K$ \cite{OrtmannAPL2008} so that at room temperature the
transport is mainly dominated by the incoherent contribution.

We have to stress that the above mentioned \textit{ad hoc} regularization of the coherent 
transport contribution will inevitably affect the quantitative 
results for the mobility. A more sophisticated treatment  which porperly incorporates the missing 
scattering effects in the case of the Holstein model and can be found in Ref. \cite{rhopolaron}, 
showing that the mobility is finite without the need to introduce extra phenomenological terms in the Hamiltonian. 
To the best of our knowledge, an analogous treatment for the Peierls-type coupling of Eq. (\ref{eq:SSH}) hasn't been developed yet.

In addition to the above mentioned \textit{ad hoc} regularization procedure, 
there is another fundamental issue which is often neglected in the application of
theories based on the polaron transformation,  and it is related to the 
 Holstein band narrowing 
 approximation Eq. (\ref{eq:bandnarrowing}). Studies have been performed 
 to assess the validity of such approximation scheme,
both at zero temperature via numerical Monte Carlo calculations combined with an analytical
expansion in powers of $1/\lambda$,
\cite{AlexKornilovicPRL1999}
 and at finite temperature 
using a generalization of the so-called momentum-average approximation \cite{BerciuPRL2006} 
to non-translationally
invariant systems. \cite{orgarpes}
Such studies have shown
that, as was originally devised by Holstein, 
the band-narrowing approximation can only be
applied  if the vibrational frequencies are much larger that the
unrenormalized electron bandwidth (the so called anti-adiabatic limit, $J\ll \hbar\omega_0$), 
because only in this case  the phonon cloud can instantaneously re-arrange to follow the motion of the carriers
as encoded in Eq. (\ref{eq:bandnarrowing}).
Unfortunately this is not the case for the inter-molecular
vibrations relevant in organic semiconductors, \cite{GirlandoPRB10,CoropceanuChemRev2007,DaSilvaAdvMat2005,HannewaldBobbertAPL2004,HannewaldPRB2004,KatoJCP2002,
SanchezCarrera-JACS10} which lie in the opposite adiabatic regime 
where the electrons can follow instantaneously the slow motion of the
molecules. Even in this case, it has been shown    that the 
intermolecular coupling is not sufficient to form an inter-molecular (bond-centered)
polaron. \cite{VukmirovicPRL2012,CSGSSHPRB1997,DeFilippisPRB2010}

We mention that ARPES experiments on Pentacene films, which reported a  bandwidth reduction upon increasing  temperature, 
were initially interpreted according  the concept of polaron band narrowing,  in apparent
 support to polaron theories for the mobility. \cite{HatchPRL2010} These measurements, however,  were subsequently
re-examined showing that the observed band narrowing  can actually be explained by accounting for the  expansion of the molecular lattice
with temperature. \cite{exporgarpes,MasinoMS2004,LiNarrowingJPCL2012}



\section{The transient localization scenario}
\label{sec:transloc}

In the preceding Sections we have listed a series of open experimental 
issues in organic semiconductors, and shown that these cannot be explained by conventional theories of charge transport.
We anticipated  that 
the cause of the  puzzling experimental observations should be seeked in the 
existence of large amplitude thermal molecular motions, which 
act as a source of  dynamical disorder for the charge carriers.
This causes a quantum localization of the wavefunctions on timescales shorter than the period of 
molecular oscillations, strongly limiting the carrier diffusion.
In this Section we describe the theoretical
framework and  numerical methods that have been developed
in recent years in order to properly address the transport properties 
in this original regime of \textit{transient localization} that is characteristic of 
organic semiconductors.

\subsection{Reconciling the band-like / localized carriers duality.}
\label{sec:reconcile}

We have seen in  Section \ref{sec:semic} that, within semi-classical transport theory, 
two very different transport mechanisms are realized
in the  limits where the fluctuations $s$ of the inter-molecular transfer integrals
are either much smaller or much larger than the equilibrium value $J$ characterizing the perfect crystal.
In the first case, transport is dominated by the diffusion of extended carriers with well-defined momentum, leading to 
Eq. (\ref{eq:loT}). In the second case, itinerant carriers are washed out and charge transport occurs
via incoherent states diffusing from site to site, leading to Eq. (\ref{eq:musat}).

These two apparently contradictory views can be reconciled by properly addressing the properties of the electronic
states as a function of their  energy and momentum.
This was done in Refs. \cite{reconcile09} and \cite{PRB12} by solving exactly 
the model Eq. (\ref{eq:SSH}) in the limit of static molecular displacements.
This limit, where thermal fluctuations are  treated as a statistical disorder in the inter-molecular bonds, 
provides a very accurate description of the excitation spectrum and carrier lifetimes in the low vibrational 
frequency regime appropriate 
to organic semiconductors. More importantly, it gives fundamental informations on 
the localization of the electronic states.

The results obtained in Refs. \cite{PRB12,reconcile09} have shown 
that at intermediate values of the ratio $s/J$, corresponding to the experimentally relevant temperature range,
both extended "band-like" carriers and incoherent excitations coexist {\it in different  regions of the excitation spectrum}: 
carriers with a markedly itinerant character are mostly located within the bulk of the band, while 
the most dramatic effects of inter-molecular fluctuations are instead concentrated around the band tails, where they cause 
the states to have a more localized  character. 
The origin of the long-standing controversy on
the microscopic identity of the charge carriers in organic semiconductors comes from the fact that
different experimental probes will see alternatively one feature
or the other, or a mixture of both.

\begin{figure}[th!]\center
\includegraphics[width=8.0cm]{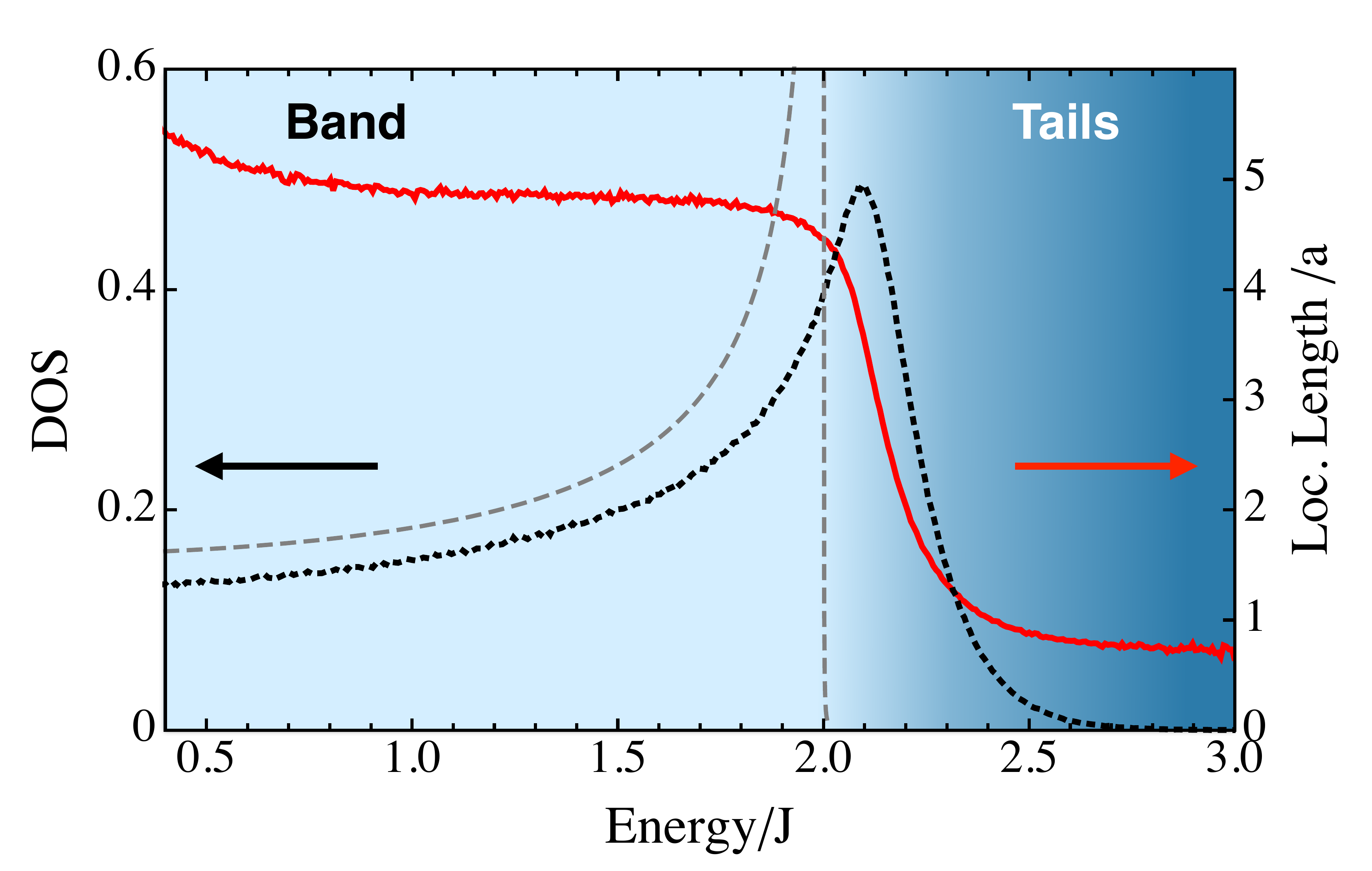}
\caption{Density of states (black dotted, left axis) and energy-resolved localization length 
in units of the lattice spacing (red, right axis) DOS  obtained 
by exact diagonalization of Eq. (\ref{eq:SSH}) in the limit of static displacements. The data correspond to the top of the HOMO band 
and are calculated for a coupling constant $\lambda=0.17$ at $T=0.3J$ (from Ref. \cite{PRB12}). 
Tails of localized states with spread $\approx a$ emerge 
beyond the band edge, whose width is controlled by the temperature via the parameter $s=\sqrt{4\lambda Jk_BT}$.
This parameter also controls the spatial extension of the band states, which rapidly decreases with temperature but
remains larger than the lattice spacing throughout the experimentally relevant temperature regime.}
\label{fig:reconcile}
\end{figure}

\paragraph{The incipient localization length.--}
Fig. \ref{fig:reconcile} reports two quantities which illustrate the duality between itinerant and incoherent 
carriers.
The dotted line is the   density of states (DOS) $\rho(\nu)$ 
at the top of the HOMO band obtained from the static solution of Eq. (\ref{eq:SSH}) (number of states per unit volume and 
unit energy $J$).
The DOS of a perfectly ordered one-dimensional crystal is shown for reference (dashed). 
Including the thermal inter-molecular fluctuations  has two visible effects.
First, the edge of the  band at
$\nu=2J$ is rounded off and shifts to higher energy, indicating an increase of
the effective bandwidth. This follows from the fact that  $\langle J_{ij}^2 \rangle >J^2$, as reported after Eq. (\ref{eq:Kubo}). 
Second, a tail of new states is generated beyond the
range of band states.
Both effects 
are controlled by the amount of inter-molecular fluctuations,  quantified through
the spread $s=\sqrt{4\lambda J k_BT}$ introduced previously. In particular, 
the width of the tails is directly proportional to $s$ and therefore increases with temperature.

The crucial information on the localized character of the states comes from the red curve in 
Fig. \ref{fig:reconcile}, which 
illustrates the spread  of the electronic wavefunctions as a function of their energy 
in the excitation spectrum.
Such quantity was introduced in 
Ref. \cite{reconcile09} in order to rationalize the idea of an underlying finite 
localization length for electrons proposed in Ref. \cite{TroisiPRL06} (see  Section \ref{sec:Ehrenfest}),
and is now commonly used in  model theoretical studies of organic semiconductors,
see e.g. Refs. \cite{Chang-Troisi-SirringhausPRL2011,PRB12,TroisiJCP2011,Li-Coropceanu-PRB12}.
The idea is that the system in the presence of slow molecular motions (where the charge carriers are mobile
as clearly shown by experiments) 
retains some important characteristics of the static disorder problem, where all 
electronic states would instead be localized. 
In particular, the  localization length calculated in the static disorder problem 
is reflected as an "incipient" localization length in the dynamical case, \cite{PRB12,PRB14,PRB11}
in a sense that will be rigorously defined in  Sec. \ref{sec:RTA}.

The comparison with the DOS allows us to identify two distinct
regions in the electronic spectrum, separated by a crossover region of width $\simeq s$
around the band edge. States located in the bulk of the band extend over many inter-molecular distances 
at low temperatures
(their spatial extent is progressively reduced upon increasing the thermal disorder, also controlled by $s$).
Tail states induced by disorder beyond the band edge instead have a much more local character, residing essentially on
one molecular unit.
The relative importance of these two 
types of excitations is  subtly controlled by the temperature: on one hand,
the amount of thermal molecular motions  controls the emergence 
of tail states and the progressive destruction  of quasi-particle band states; on the other hand,
the temperature also controls the relative weight of these states, via their
statistical population in the respective energy ranges. \cite{PRB12}

\paragraph{Kubo formula in the bubble approximation.--}
While the results presented above provide an essentially exact description of the properties of the electronic states, 
more input is required in order to assess how these single-particle states participate in the transport mechanism, which involves
a two-particle correlation function.  \cite{Kubo57,Mahan}
To this aim, an approximate treatment 
was developed in Refs. \cite{reconcile09} and \cite{PRB12}, which starts from the  exact numerical
calculation of the Green's function in the static disorder limit  and 
assumes a convolution form for the current-current correlation function 
in the Kubo formula
as in Eq. (\ref{eq:Kubo}). 
The corresponding mobility can be written in compact form as \cite{rhopolaron,PRB12,reconcile09}
\begin{equation}
  \label{eq:mobzero}
  \mu=\frac{e}{k_BT}\frac{\int d\nu
    e^{-\beta \nu}B(\nu)}{\int d\nu \rho(\nu)e^{-\beta \nu}},
\end{equation}
where $\rho(\nu)=\mathrm{tr} \hat \rho (\nu)$ is the interacting DOS with  
$\hat \rho (\nu)= -Im (\nu-\hat{H})^{-1}/\pi$.
The function $B(\nu)$, 
which is proportional to  an energy-resolved diffusivity,
is defined  in terms of the current operator $\hat{J}$ as
\begin{equation}
B(\nu)= \frac{\pi}{ e}\mathrm{tr}[\left\langle \langle  \hat \rho (\nu) \rangle \hat J \langle 
 \hat\rho (\nu) \rangle \hat J \right\rangle]
\end{equation}
where the trace is performed over lattice sites and angular brackets represent averages over the classical phonon field.
This approach is  more general than the one presented in 
Sec. \ref{sec:incoherent}, as it includes the full momentum and energy dependence of the electronic spectral 
function. 
In particular,  it is  able to describe the simultaneous presence of 
both band and incoherent states, acting as two complementary transport channels.

The results for the mobility obtained by this method are reported in Fig. \ref{fig:semic}  (full line). 
The calculated mobility interpolates smoothly between the 
two semi-classical limiting behaviors presented in Section \ref{sec:semic}:
the transport is essentially governed by band-like carriers at low temperature, i.e. when thermal molecular disorder is small, 
corresponding to Eq. (\ref{eq:loT}); 
upon increasing the temperature, the relative weight of the incoherent states
progressively increases up to the point where band states get completely washed
out, as their lifetime becomes shorter than the inter-molecular transfer time. The mobility
then tends to the fully incoherent expression Eq. (\ref{eq:musat}).

Importantly, Fig. \ref{fig:semic} shows that  the present treatment, albeit more sophisticated
than the ones of Section \ref{sec:semic}, 
is still unable to restore a quantitative agreement with the experiments. This failure in reproducing the
experimental results indicates that
localization effects in organic semiconductors are  stronger than what can be included
in any extension of semi-classical theory.

\begin{figure}[th!]\center
\includegraphics[width=7.6cm]{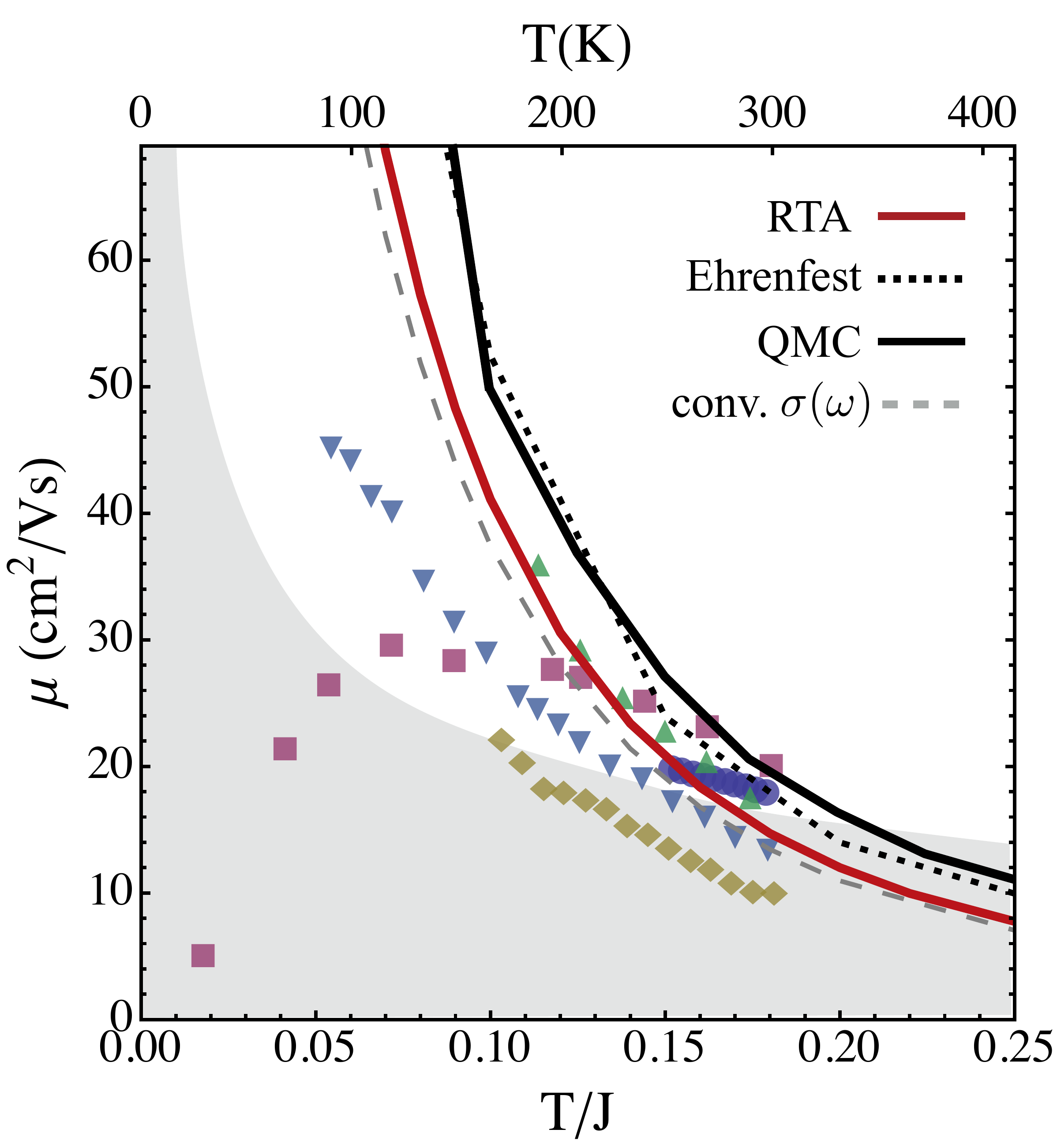}
\includegraphics[width=7.7cm]{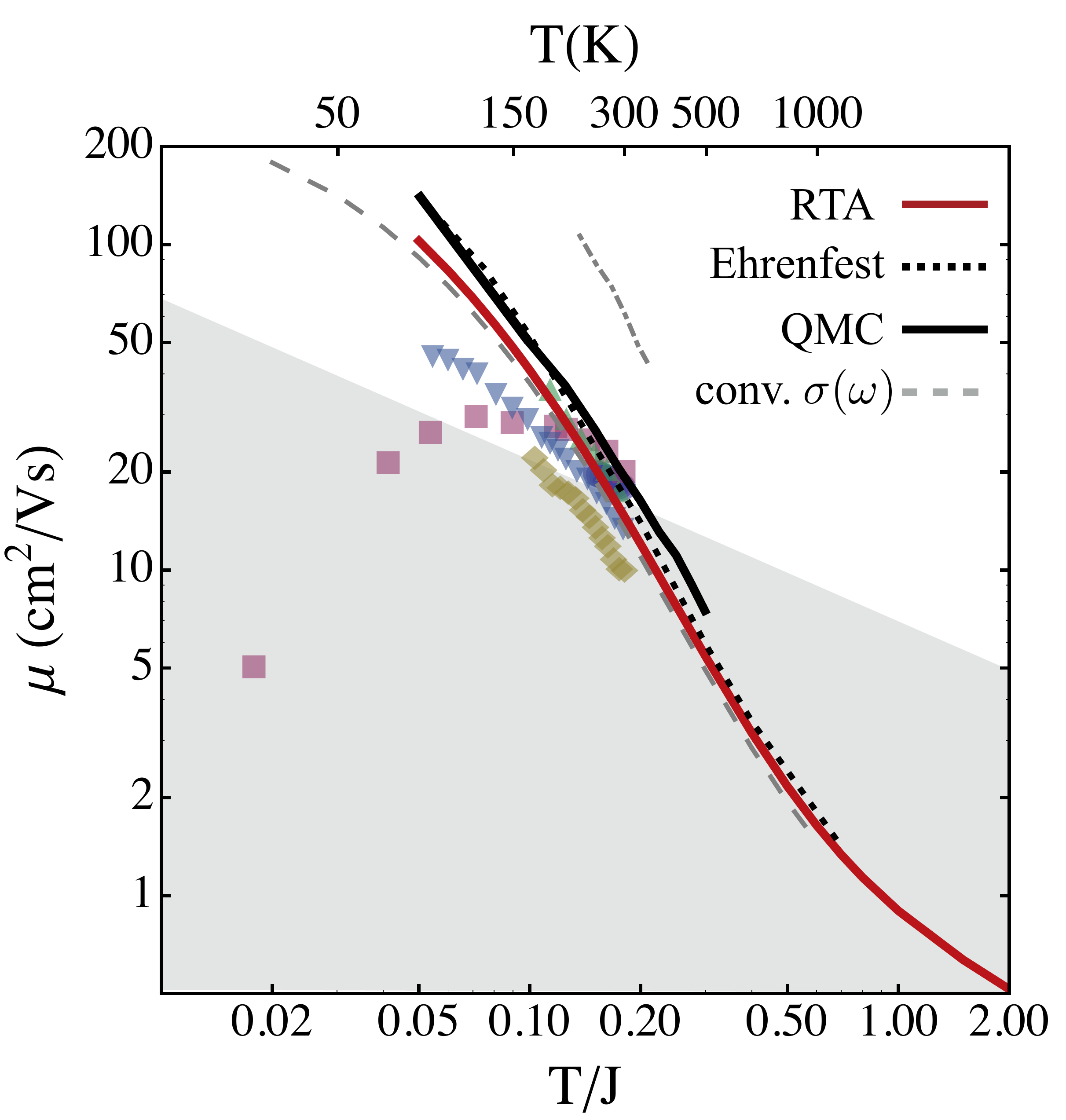}
\caption{
Top: Mobility calculated from Eq. (\ref{eq:SSH}) using theories which account 
for quantum localization processes: RTA, Ehrenfest simulations, QMC and 
Lorentzian convolution of the optical conductivity (see text). 
The symbols for the experimental data are the same as in Fig. \ref{fig:semic}.
The gray shaded area shows the Mott-Ioffe-Regel limit defined by Eq. (\ref{eq:muMIR}).
Bottom: same data, plotted in log scale, for a direct comparison with Fig. \ref{fig:semic}.
The dash-dotted line is the Ehrenfest result of Ref. \cite{TroisiAdvMat2007}.
}
\label{fig:mobility}
\end{figure}

\subsection{Ehrenfest simulations}
\label{sec:Ehrenfest}

The pioneering work of Troisi and Orlandi in 2006 \cite{TroisiPRL06} suggested the possibility
that the carriers in organic semiconductors undergo some form of localization due to the strong scattering introduced by the large molecular motions. 
These authors started from the idea that because molecular motions are slow, 
they can be treated classically and separated from the faster electronic motion in a Born-Oppenheimer scheme. 
Accordingly, they performed numerical simulations where the electronic problem  in Eq. (\ref{eq:SSH}) 
is solved exactly at each instant of time following the classical evolution 
of the slow degrees of freedom $\lbrace u_i \rbrace$, through the Ehrenfest method.
We summarize this method here, following the description given in Ref. \cite{PRB11}.

\begin{figure}[th!]\center
\includegraphics[width=8cm]{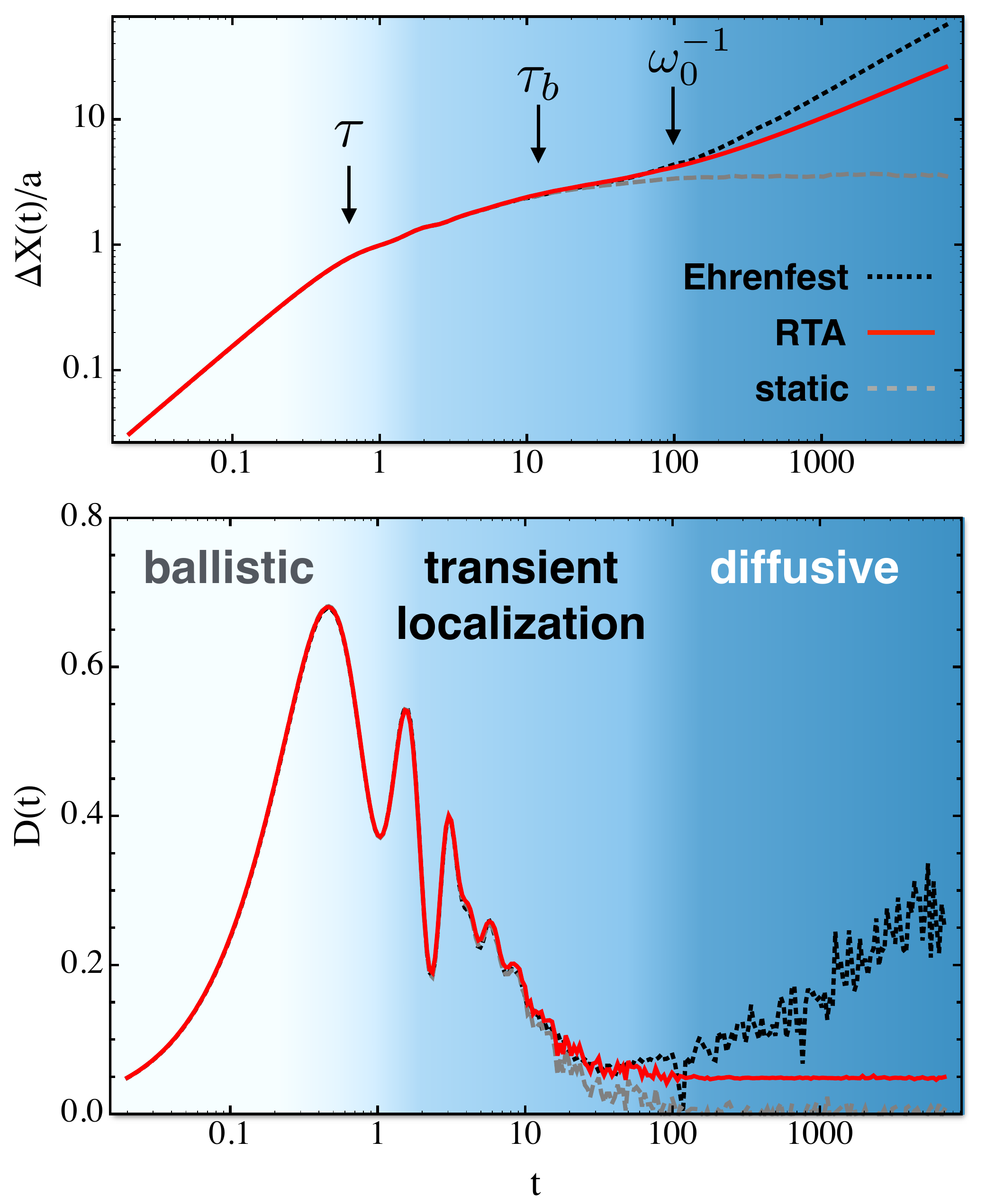}
\caption{Top: time-dependent quantum spread of the electronic wavefunction, $\sqrt{\Delta X^2(t)}$, calculated for the model 
Eq. (\ref{eq:SSH}) via the 
Ehrenfest method (black dotted curve) and via the RTA by setting $\tau_{in}=\omega_0^{-1}$ (red curve). The result for static
molecular displacements is shown for comparison (gray dashed).
Bottom: the corresponding instantaneous diffusivity $D(t)=d\Delta X^2(t)/2dt$.
The parameters are $J=110meV$, $\lambda=0.25$, $\omega_0=0.01J$, $T/J=0.2$ (readapted from Ref. \cite{PRB11}). 
Time is expressed in units of $\hbar/J$.} 
\label{fig:Ehrenfest}
\end{figure}

The coupled equations of motion for the electrons and molecular degrees of freedom read:
\begin{eqnarray}
& &   i\hbar \partial_t |\Psi,t \rangle = H_{el}(\{ u_i \}) |\Psi,t\rangle
\label{eq:Scrod} \\
& &  \ddot{u}_i = -\omega^2_0 u_i -\frac{\partial}{\partial u_i} \langle t,\Psi| H_{el}(\{ u_i \})|\Psi,t \rangle,
\label{eq:Ehrenfest}
\end{eqnarray}
where $H_{el}$ is the electronic part of Eq. (\ref{eq:SSH}).
In the Ehrenfest method, the transport properties are calculated by tracking the
quantum spread of the electronic wavefunction  as the mean-square displacement  $\Delta X^2(t)=\langle \Psi |[\hat X(t)-\hat X(0)]^2 |\Psi \rangle$ in real time, 
as illustrated in Fig. \ref{fig:Ehrenfest}(a) (readapted from 
\cite{PRB11}).  This is done by following simultaneously the evolution of the state vector
$|\Psi,t\rangle$ and that of  $|x\Psi,t\rangle$,  which is the evolution of the 
 state $\hat X|\Psi,0\rangle$. The quantum spread is then given by
$\Delta X^2(t)=\langle t,x\Psi|x\Psi,t\rangle + \langle 0,x\Psi|x\Psi,0\rangle- 2 Re 
\langle t,x\Psi|x\Psi,0\rangle$.
The time derivative  of the spread directly provides the instantaneous diffusivity
$D(t)=d\Delta X^2(t)/2dt$ (Fig. \ref{fig:Ehrenfest}(b)). The mobility is then extracted 
from the diffusivity  in the long time limit $D=\lim_{t\to \infty} D(t)$ through the Einstein formula, 
$\mu=eD/k_BT$. \cite{Kubo57} 

Finite temperature simulations are performed starting from a thermal state of
the classical oscillators and taking the initial state of the electron as an
eigenstate of  $H_{el}$ in the corresponding distorted landscape, with a probability
proportional to the Boltzmann weight. Since the initial oscillator state has no polaronic disortion in it, the 
(small) initial correlations between the charge and the lattice degrees of freedom 
are lost in the dynamical evolution.
Starting instead from a localized electron-lattice correlated polaronic initial state,
these correlations are preserved at least at low temperature by the 
Ehrenfest dynamics. \cite{MozafariStafstromJCP2013} This case, however, 
does not apply here: for the model Eq. 
(\ref{eq:SSH}) a polaronic ground state can only be found  for $\lambda \ge 0.5$,
\cite{CSGSSHPRB1997,DeFilippisPRB2010} which is well above the values typical for organic crystals.

\paragraph{Beyond semi-classical transport.--}
The key advantage of the Ehrenfest method over previous approaches of charge transport in organic semiconductors is that 
it is  able to account for quantum localization
corrections in the sense of Anderson, because the electronic degrees of freedom are treated exactly.  
In other words, these mixed quantum-classical simulations are not "semi-classical" from the point of view of 
electronic transport. For this reason, and because they provide a direct way of visualizing the dynamics
of charge carriers, Ehrenfest simulations  have been extensively used in recent years both for one-dimensional
and two-dimensional models for organic semiconductors.
\cite{TroisiPRL06,StafstromChemSocRev2010,PRB12,TroisiAdvMat2007,TroisiJCP2011,PRB11,MozafariStafstromJCP2013,
WangPCCP2010,Wang-BeljonneJCP2011,Ishii-MobilityWavePacket-PRB2012,Tamura-Eherenfest2d-PRB2012,
Ishii-Ehrenfest-2D-anisotropy-PRB13,Ishii-Hall-PRB2014}


A close look at  Fig. \ref{fig:Ehrenfest}(a)
shows that the behavior of the time-dependent electronic spread calculated 
with the Ehrenfest method is quite different from what would be expected
from semi-classical diffusion. In semi-classical transport, the 
electronic wavefunction spreads ballistically at initial times, as $\Delta X^2(t)=\langle V^2\rangle t^2$, up to 
some characteristic scattering time $\tau$ where a diffusive behavior, $\Delta X^2(t)=(D_{sc}/2)t$, sets in 
(this behavior corresponds to the dashed line in Fig. \ref{fig:phenom} below).
\cite{PRB12,PRB14,PRB11}
Correspondingly the instantaneous 
diffusivity $D(t)$  would increase linearly at short times and then saturate to its
long time limit $D_{sc}$ at $t\gg \tau$. 
What is observed instead is that, after the initial ballistic behavior, the mean-square displacement 
of the electron  
bends down and tends to saturate at a time   $\tau_b> \tau$, 
before diffusing again at a subsequent time $\tau_{in}$. 
As a result, the diffusivity goes through a maximum 
and then decreases indicating the onset of 
localization. This decrease goes on up to the time $t \sim \tau_{in}$. The resulting 
diffusivity is considerably lower than the semi-classical value $D_{sc}$, which roughly corresponds to
the maximum of $D(t)$. 
This peculiar non-monotonic behavior of the diffusivity 
identifies the transient localization mechanism for charge transport.

 Comparing the numerical results at different values of the inter-molecular 
vibration frequencies shows that the time where the diffusion is restored is directly proportional 
to the  timescale of molecular motions,  
$\tau_{in} \sim 1/\omega_0$. \cite{PRB11} 
The behavior 
at times $t\lesssim \tau_{in}$ is indistinguishable from what would be obtained
in the presence of static molecular displacements (dashed curve in Fig. \ref{fig:Ehrenfest}(a)),
 because the molecular motions effectively appear as frozen at such short time-scales.
The behavior of the dynamically disordered system
therefore "knows" about the existence of quantum localization processes 
and of a finite localization length for the electrons,
even though a diffusive behavior is eventually obtained at times $t\gtrsim \tau_{in}$.

\paragraph{Drawbacks.--}
Despite its popularity, the Ehrenfest  method suffers  from 
a number of drawbacks.
The first originates from the fact that the back-action of the electrons  on the 
molecular motions is treated at mean-field level, in the form of an average instantaneous force (cf. Eq. (\ref{eq:Ehrenfest})).
It is known \cite{Beljonne-SurfaceHopping-JPCLett2013,PRB11,ParandekarJCP2005} 
that this back-action is not sufficient to  
thermally equilibrate the system, and results in a progressive heating of the electronic 
system due to the excess energy being constantly injected by the molecular vibrations. 
The consequences are visible in Fig. \ref{fig:Ehrenfest}(b) (black dotted line): the diffusivity is not constant at long times,
but increases steadily due to heating, because more and more conducting states 
within the band are artificially populated. This can lead to an over-estimate of the mobility that 
spuriously depends on 
the simulation time, which could be at the origin of the discrepancy between the results of 
\cite{TroisiAdvMat2007} and those obtained by other authors using the same Ehrenfest method. To illustrate this point, 
in Fig. \ref{fig:mobility} we compare the data of Ref. \cite{TroisiAdvMat2007} (dash-dotted line) with
the data obtained by the same method and for the same set of parameters in Ref. \cite{PRB12} (dotted line), 
but with the prescription of evaluating the 
diffusivity at a fixed time $t=1/\omega_0$: the mobility is much lower  in the latter case,
and is extremely close to the fully quantum calculation of Ref. \cite{KataQMC} (see Sec. \ref{sec:QMC} below).

Also due to the average nature of the back-action term, 
the Ehrenfest method does not give access to the thermally activated mobility of 
finite-radius polaronic states. For polarons to be formed (and stable in a dynamical evolution) a feedback to
the oscillator motion which goes beyond the average Ehrenfest method is needed.
Starting from a localized electron-lattice correlated polaronic initial state 
a small-polaron evolution can be followed for  limited times 
 in the Ehrenfest dynamics. \cite{MozafariStafstromJCP2013}
However 
the averaged oscillator dynamics of the Ehrenfest method cannot sustain polaronic correlations
in the long time limit  even at large values of the electron-phonon coupling, unless the temperature is sufficiently small. 
This drawback was corrected by Wang and Beljonne in Ref. \cite{Beljonne-SurfaceHopping-JPCLett2013} by introducing 
surface hopping methods which go beyond the
mean field character of the Ehrenfest average force (see Sec. \ref{sec:other}).

The third issue is
that wavefunction-based simulations cannot distinguish between 
the evolution of a state of finite radius whose center of mass diffuses over time (as in a classical random walk) and an 
evolution where the wavefunction itself spreads over time, 
unless the electron-phonon interaction has a large local contribution as in Ref. \cite{MozafariStafstromJCP2013}.
For the electron-phonon couplings typical of organics crystals, it is rather the
spread $\Delta X^2$ of the wavefunction {\it for a single initial set} of molecular
displacements which grows with time, leading to a fully delocalized wavefunction
in the long time limit (see Ref. \cite{YaoYao-JCP12} for an approach which includes
decoherence in order to solve this issue). The interpretation of the time snapshots 
of the electronic density should therefore taken with care.


Finally we emphasize that Ehrenfest based methods are not appropriate to treat 
the effects of high frequency intramolecular vibrations. When they are used 
in this regime, they invariably lead to  unrealistically low mobilities, 
in contrast with the results of fully quantum treatments which predict 
only a moderate reduction of the mobility due to polaronic band renormalization 
(cf. Section \ref{sec:hopping}).

\subsection{Relaxation time approximation}
\label{sec:RTA}

The existence of a direct connection with the static disorder problem, that was anticipated 
in Sec. \ref{sec:reconcile} and
demonstrated by the Ehrenfest simulations (cf. Fig. \ref{fig:Ehrenfest}(a)),
suggests that despite the apparent band-like behavior of the mobility,
the charge transport mechanism in organic semiconductors could be better understood
by adopting a radically different paradigm, 
which takes localization  as a starting point. 
A simple scheme to bridge between static and dynamical disorder has been developed in Refs. 
\cite{PRB12,PRB14,PRB11} based on a relaxation time approximation (RTA) applied to the localized limit. \cite{MayouPRL2000,TramblyPRL2006}
In addition to providing a very efficient method for the calculation of the mobility,
which overcomes the drawbacks of the Ehrenfest method,
the RTA scheme has the advantage of providing a transparent analytical insight 
on the microscopic mechanism of charge transport,  clarifying both 
the relationship between mobility and localization and 
the crucial effects of molecular dynamics. 
Recently, it has been shown that the RTA becomes exact in a model for 
exciton transport where the inelastic scattering time is replaced by the decoherence time of the exciton.
 \cite{MoixNJP2013}
The effects of short-time correlations in the dynamics of disorder have been recently 
studied in Ref.  \cite{Packwood2015} using a stochastic model similar to Ref. \cite{MadhukarPostPRL1977}.

\paragraph{Analytical insights on the transport mechanism: \\ transient localization length and inelastic scattering time.--}
The idea underlying the RTA is to express the dynamical properties of the system under study  in 
terms of those of a suitably defined reference system, from which it decays over time.
\cite{PRB12,PRB11} As suggested
from the preceding discussion, our reference
system of choice will be an idealized version of the organic semiconductor where the molecular 
displacements are frozen. Such a reference system 
displays Anderson localization of the carriers. \cite{LeeRMP1985}

The key physical quantity for the RTA is the velocity {\it anticommutator} 
correlation function $C(t)=\lbrace \hat V(t),\hat V(0)\rbrace$, 
which is proportional to the time derivative of the
instantaneous diffusivity,  $C(t)=2dD(t)/dt$ \cite{PRB12,PRB14,PRB11}.
Taking for reference  the velocity correlation function $C_0(t)$ of 
a system with static molecular displacements, we introduce the relaxation time approximation as follows:
\footnote{Note that the standard description of semi-classical transport can be obtained by defining a reference 
$C_0$ corresponding to the opposite limit of a perfectly periodic crystal, and subsequently 
including the scattering of Bloch waves by phonons and impurities via an analogous exponential 
relaxation term. \cite{DresselGrunerBook}}
\begin{equation}
\label{eq:RTA}
C_{RTA}(t)=C_0(t)e^{-t/\tau_{in}}.
\end{equation}
It is clear from the above equation that the correlation function $C_{RTA}(t)$ 
 coincides with that of the system with static disorder  for
$t\ll \tau_{in}$, because in this case $e^{-t/\tau_{in}}\simeq 1$. 
At longer times, however, the exponential term in Eq. (\ref{eq:RTA}) causes a decay of 
the velocity correlations. This decay physically corresponds to the destruction of the quantum interference
processes that are at the origin of Anderson localization (the so-called backscattering terms), \cite{ThoulessPRL1977,LeeRMP1985} 
and that are encoded in the reference $C_0(t)$. According to the discussion in the previous Section, 
one should set $\tau_{in}$ of the order of the timescale of molecular motions, $1/\omega_0$.
\begin{figure}[th!]\center
\includegraphics[width=8cm]{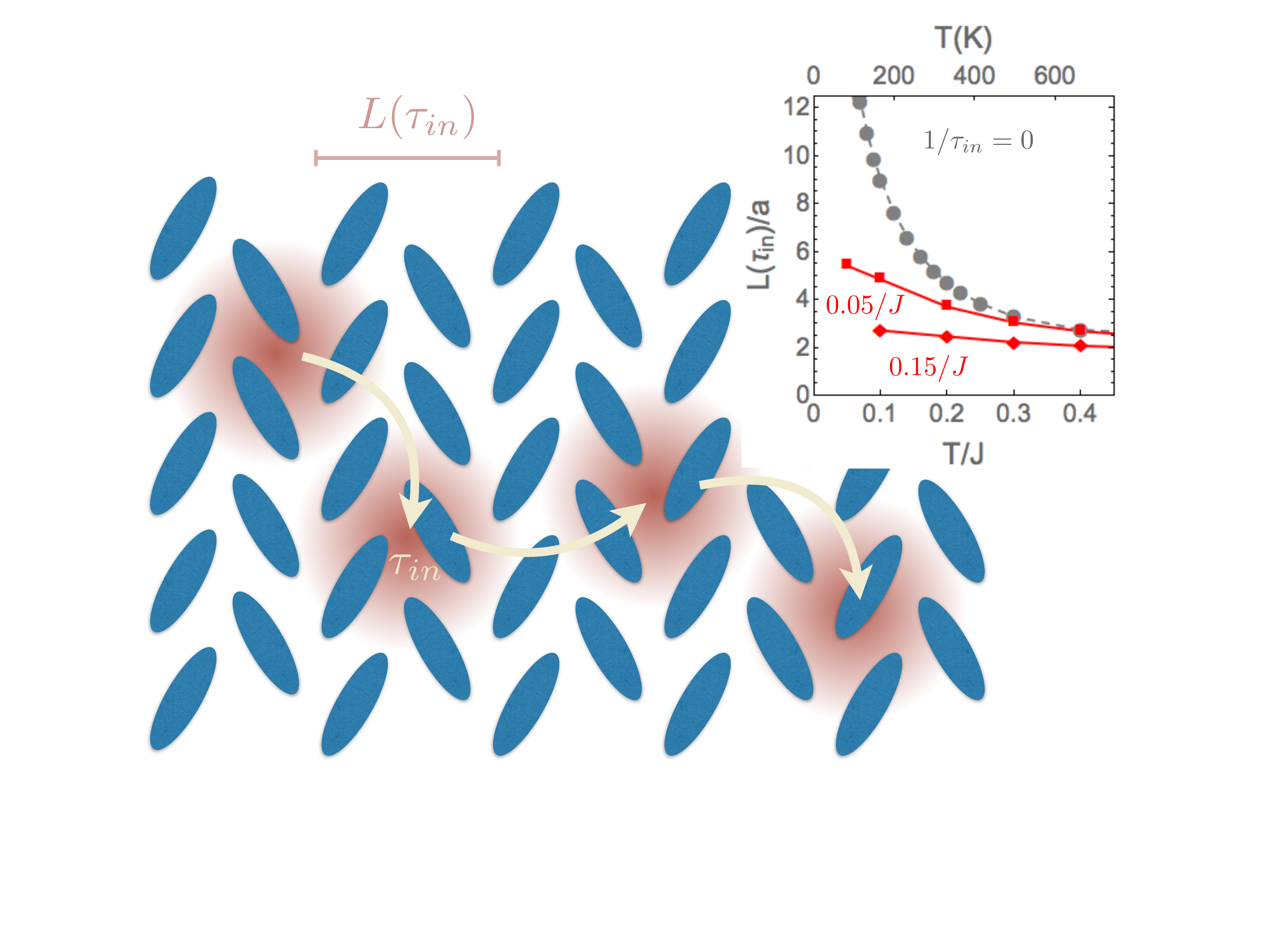}
\caption{A sketch of the transport mechanism in the transient localization regime.
The inset shows the temperature dependence of 
transient localization length $L(\tau_{in})$ of Eq. (\ref{eq:DRTA}) for two 
different values of the inelastic scattering time.}
\label{fig:sketch}
\end{figure}
The spread $\Delta X_{RTA}^2(t)$ and instantaneous diffusivity $D_{RTA}(t)$ can be readily obtained
via time integration of Eq. (\ref{eq:RTA}), and are illustrated in Fig. 
\ref{fig:Ehrenfest} (red thick lines). They exhibit the same qualitative behavior as seen in the Ehrenfest simulations, 
but do not suffer from the heating problem at long times.

From Eq. (\ref{eq:RTA}), the diffusion constant in the long time limit can be expressed as
\begin{equation}
\label{eq:DRTA}
D_{RTA}=\frac{L^2(\tau_{in})}{2\tau_{in}}.
\end{equation} 
Here $L^2(\tau_{in}) =\int e^{-t/\tau_{in}}
\Delta X^2_0(t)dt/\tau_{in}$ is the
electron spread achieved at a time $t\approx\tau_{in} \equiv 1/\omega_0$, just before
diffusion sets back in (cf. Fig. \ref{fig:Ehrenfest}). $L(\tau_{in})$
 therefore  
has the meaning of a {\it transient localization length}. The slower the molecular motions, 
the closer it approaches the actual localization length $L_0$ of the reference system, 
which is exactly attained in the limit where $\tau_{in} \to \infty$:
$L(\tau_{in}\to \infty)=L_0$.  The temperature dependence of both the static and transient localization length
is illustrated in the inset of Fig. \ref{fig:sketch}. 

The expression Eq. (\ref{eq:DRTA}) is analogous to the
Thouless diffusivity of Anderson insulators. \cite{ThoulessPRL1977} It represents
a physical process where localized electrons diffuse over a distance 
$L(\tau_{in})$ with a trial rate $1/\tau_{in}$, as pictorially illustrated in Fig. \ref{fig:sketch}. 
The corresponding mobility can be obtained from the Einstein
relation as
\begin{equation}
   \label{eq:muRTA}
   \mu_{RTA}= \frac{e}{k_BT}\frac{L^2(\tau_{in})}{2\tau_{in}}.
\end{equation}
The  mobility calculated via Eq. (\ref{eq:muRTA}) by taking 
$\tau_{in}\omega_0=1$ and the microscopic parameters appropriate for rubrene is reported 
in Fig. \ref{fig:mobility} (red full line). Several remarks are in order:

(i) From Eq. (\ref{eq:muRTA}) we can  understand the origin of the 
"metallic-like" power-law behavior of the mobility in organic semiconductors, as arising both from  
the explicit $1/T$ factor and from the temperature dependence of the
transient localization length  $L(\tau_{in})$. The latter decreases with temperature
because the carriers tend to be more and more localized upon increasing the thermal 
molecular disorder, as shown  in the inset of Fig. \ref{fig:sketch}.
In the  regime of parameters relevant to rubrene, the RTA results for the model Eq. (\ref{eq:SSH}) are well
reproduced by the following functional form:  \cite{PRB12} 
\begin{equation}
\label{eq:L2}
L^2(\tau_{in})\simeq   a^2\frac{c_{\tau_{in}}}{\lambda^l} \left(\frac{J}{T}\right)^q.
\end{equation}
with a prefactor $c_{\tau_{in}}$ which depends weakly on $\tau_{in}$.
Fitting the numerical data for $1/\tau_{in}=0.05J$ 
with the above form in the interval $0.14<\lambda<0.21$ and 
$0.18<T/J<0.3$ yields $l= 1.64\pm 0.03$ and $q= 0.94\pm 0.02$. 
Substituting this into Eq. (\ref{eq:muRTA}) leads to a mobility varying as 
$\mu\propto T^{-p}$ with $p=1+q$, i.e. roughly 
\begin{equation}
\label{eq:powerlaw}
\mu\propto T^{-2} 
\end{equation}
(see Ref. \cite{PiconPRB2007} for similar arguments applied to the case of acoustic phonons).

(ii) The temperature dependence predicted above corresponds to a system kept at constant volume. It is known
however that the lattice parameters in  organic semiconductors change appreciably with temperature, which is accompanied
by changes in the transfer integrals \cite{exporgarpes,MasinoMS2004,LiNarrowingJPCL2012}. Obviously, 
such changes
can also affect the behavior of the mobility.

(iii) As can be seen from the comparison of Figs. \ref{fig:semic} and \ref{fig:mobility}, 
the mobility in the presence of transient localization 
is considerably lower than that of semi-classical carriers (cf. also the instantaneous 
diffusivity of Fig. \ref{fig:Ehrenfest}). 
It is then easy to understand that Eq. (\ref{eq:muRTA}) can properly
describe  mobilities that, as in the experiments, fall below the so-called Mott-Ioffe-Regel limit  (cf. Sec.\ref{sec:breakdown}),  which 
is where  the apparent mean-free path 
falls below the typical inter-molecular distance $a$.
We also understand from Eq. (\ref{eq:muRTA}) why 
organic semiconductors are a particularly favorable ground for this breakdown to occur:
the large thermal molecular disorder leads to short $L(\tau_{in})$ (which reduce 
to few lattice spacings  at room temperature even in pure samples) and the 
 large values of the molecular mass lead to a large $\tau_{in}$, 
both effects contributing to produce low values of $\mu$ in Eq. (\ref{eq:muRTA}). 
Fig. \ref{fig:mobility} illustrates that the Mott-Ioffe-Regel limit is indeed approached 
in pure samples around room temperature, both in the experimental and in the theoretical results.

(iv) It is clear from Fig. \ref{fig:Ehrenfest} that the transient localization phenomenon
only occurs provided that the inelastic scattering time is longer than the time $\tau_b$
that characterizes the onset of localization, 
$\tau_{in}\gtrsim\tau_b$, or equivalently,
$\omega_0 \lesssim \hbar/\tau_b$ (typical values of $\hbar/\tau_b$ range from few units to few tens 
of meV, see Sec. \ref{sec:phenom} and Ref. \cite{PRB14}). 
In the opposite regime, i.e. for sufficiently large values of $\omega_0$, 
localization processes are completely washed out and semi-classical transport is recovered. 
Conversely, if the inter-molecular transfer integrals are too low or if the electron-vibration coupling is larger than 
a critical value $\lambda_c=0.5$, the electronic wavefunction becomes 
self-trapped to essentially a single inter-molecular bond,
$L \simeq a$, and one can expect the transport mechanism to become thermally activated.

(v) Finally, we mention that the RTA results for the mobility
 are very close to the results of both the Ehrenfest simulations and the fully quantum simulations of Ref. 
\cite{KataQMC} (see Sec. \ref{sec:QMC}).
\footnote{A detailed (unpublished) study of the take-off time of the quantum-spread in the 
Ehrenfest simulations suggests that these are closely described by the RTA if one sets 
$\tau_{in}\omega_0=(0.5 \pm 0.1)$.  Taking this value in Eq. (\ref{eq:muRTA})  
would slightly increase the mobility compared to what is shown in Fig. \ref{fig:mobility},
improving further the agreement with the QMC results.}
The relatively modest computational cost of the RTA  makes this method very interesting
to numerically access large system sizes, \cite{Mayougraphene13} 
or to perform systematic screening studies of different compounds.

\subsubsection{Strategies to improve the mobility}
Beyond its numerical versatility, 
the analytical content of the RTA formula Eq. (\ref{eq:muRTA}) can be used to make
systematic predictions on how the mobility varies upon changing the microscopic parameters,
in the regime where transient localization applies (i.e. for sufficiently large $J$, not too large $\omega_0$ 
and not too large $\lambda$, in the sense specified in the preceding paragraphs). 
This can provide useful strategies to improve the performances of real compounds 
(a recent work describing different engineering strategies can be found in Ref. 
\cite{Dong-highmobility-review-AdvMat13}).
The following possibilities can be explored:

\textbf{(i) Increasing the molecular overlaps, and therefore the transfer integral $J$.}
This could be achieved in principle by optimizing the crystal packing. Practical examples 
include chemical functionalization,  the application of pressure, or strain \cite{BaoNcomms2014} 
(in thin films or self-assembled monolayers).  
From the theoretical point of view, we observe that within the transient localization scenario
there is no explicit  dependence of the mobility on $J$, cf. Eq. (\ref{eq:muRTA}).
Variations of the transfer integrals will therefore only enter implicitly  through 
the behavior of the transient localization length. 

Using the analytical expression Eq. (\ref{eq:L2}) valid in the regime $L\gtrsim a$ 
for the typical microscopic parameters that apply to rubrene, 
we obtain  approximately
$$\mu \propto J^q \simeq J$$ 
at constant  $\lambda$ (because $q\simeq 1$).
It has to be recognized, however,  that any variation of the electron-vibration coupling strength 
$\lambda$ induced by structural modifications will also add to the power-law dependence (see also point (ii) below). 
Although it is not possible to assess precisely how this parameter depends 
on the crystal structure without a detailed calculation, from the behavior depicted in Fig. \ref{fig:Troisi}
one can imagine two limiting situations: \\

\noindent
\textit{Optimizing the $\pi-\pi$ distance}.  If the equilibrium perpendicular distance between the molecules is varied without
changing the long-axis molecular displacement, then the relative slope of $J_{ij}(u_i-u_j)$ measured
in units of the equilibrium $J$ does not change. By definition, this amounts to keeping the 
parameter $\alpha=(1/J)(dJ/du)$ fixed, as can be directly checked from Eq. (\ref{eq:SSH}). 
In this case, from the expression $\lambda=\alpha^2 (\hbar/2M\omega_0) (J/\hbar\omega_0)$
we see that $\lambda$ {\em increases} proportionally to $J$, which yields for the mobility 
$$\mu\propto J^{q-l}\simeq J^{-0.7}.$$ 
We conclude that, quite unexpectedly, a reduction of the 
perpendicular ($\pi-\pi$) distance between neighboring molecules along a stack will be 
detrimental to the mobility.\\

\noindent
\textit{Optimizing the long-axis distance}.  If instead the lateral coordinate is varied without changing the perpendicular distance,
then the parameter $\alpha$ changes with $J$, and the coupling constant $\lambda$ also changes accordingly. 
Adopting a linear approximation for $J_{ij}(u_i-u_j)$, 
which is generally valid far from the extrema of the curve  in Fig. \ref{fig:Troisi},
one has in this case $\alpha=(1/J)(dJ/du)\propto 1/J$. Consequently 
for long-axis displacements the coupling  decreases upon
increasing the transfer integral as $\lambda\propto 1/J$, which adds up to give 
$$\mu \propto J^{q+l}\simeq J^{2.6}.$$
Note that if the equilibrium displacements $u_i-u_j$ in a given material are already
close to the extrema (as is the case for rubrene, cf. Fig. \ref{fig:Troisi})  
neither $J$ nor $\lambda$ depend much on the equilibrium displacement $u_i-u_j$, so that
optimizing the structure will not lead to major improvements of the mobility.

The three  contradictory behaviors identified above 
imply that it is difficult to identify a general rule predicting 
how the mobility depends on the inter-molecular transfer integrals, 
and case by case structural calculations are needed.
This might explain why  recent studies 
on crystals of functionalized rubrene molecules \cite{GarryCM2013}  could not identify a clear trend relating 
$\mu$ and $J$.

\textbf{(ii) Reducing the coupling with the inter-molecular motions}, which results in 
an increase of the transient localization length. The discussion 
at point (i) indicates that this can be achieved by adjusting the crystal packing 
to minimize the dependence of the inter-molecular
transfer with inter-molecular distance, i.e. the slope of $J_{ij}(u_i-u_j)$, which is essentially the parameter 
$\alpha$ of Eq. (\ref{eq:SSH}). 
Accounting for the  dependence of the transient localization length Eq. (\ref{eq:L2}) on 
the coupling constant yields
$$\mu \propto \lambda^{-l} \propto \alpha^{-2l},$$ with $l \simeq 1.64$.

\textbf{(iii) Increasing the inter-molecular vibration frequency $\omega_0$} by tightening the inter-molecular bonds.
The frequency of inter-molecular vibrations enters explicitly in the formula Eq. (\ref{eq:muRTA})
for the mobility:
varying $\omega_0$ while keeping $\lambda$ and $J$ 
fixed should affect the mobility linearly, because $1/\tau_{in}\propto \omega_0$. 
The numerical results of \cite{PRB12} show that the effect is actually 
sub-linear, $\mu\propto \omega_0^\zeta$, with $\zeta\simeq 0.35$ for small variations 
around $\omega_0/J=0.05$.
This weaker power law  can be ascribed to the appreciable dependence of $L(\tau_{in})$ on $\tau_{in}$:
the transient localization length diminishes when $\tau_{in}$ decreases, i.e. when 
the frequency of the molecular motions  increases, cf. Figs.
\ref{fig:Ehrenfest} and \ref{fig:phenom}. 

To assess the overall effect of an increase of $\omega_0$  on the mobility,
one has again to account for the explicit dependence of the coupling to inter-molecular motions. The coupling constant
strongly decreases as
$\lambda\propto 1/\omega_0^2$, which
leads to a total 
$$\mu \propto \omega_0^{2l+\zeta} \simeq \omega_0^{3.6}.$$
Tightening the inter-molecular bonds therefore appears as a very efficient
strategy to increase the mobility in  organic semiconductors. This could be achieved for example by functionalizing the
molecules with rigidly bound side groups, which would confine the long-axis displacements of the molecules, 
effectively increasing $\omega_0$. Similarly, it is expected that smaller molecules will have less ease
in sliding along the long axis, also leading to generally larger vibrational frequencies.
We note that for sufficiently large values of $\omega_0$, 
eventually a semi-classical transport mechanism should 
recovered, again leading to appreciably higher values of the mobility as illustrated 
in Fig. \ref{fig:semic} (see the discussion 
at point (iv) in Sec. \ref{sec:RTA}).



We note that changing the molecular mass $M$ (as in an isotope substitution experiment) does not affect the 
electron-vibration coupling constant $\lambda$. This can be directly checked 
from the  expression of $\lambda$ given above, and noting that $\omega_0=\sqrt{K/M}$ with $K$ the 
force constant.  As a result, the effect of mass substitutions on the mobility would be weak.


\subsection{Other quantum approaches}
\label{sec:other}


\paragraph{Surface hopping methods.--}
 In the original
Eherenfest method, the oscillators evolve along a single, {\it averaged}
potential.
The surface-hopping method developed by Tully 
\cite{TullySurfaceHopping} is a dynamical procedure which modifies Ehrenfest's
equations of motion for classical oscillators to include the switching from
one adiabatic surface to another, 
  which is valid under the assumption that electronic coherence is lost 
in a time shorter than the average surface hopping time.
This allows to take into account correlations between the lattice displacements and the electron
density, giving access to processes close to the small polaron incoherent hopping regime 
(the Ehrenfest method can, on the contrary, be considered too coherent and the chemical physics 
literature contains methods to reduce its coherence time \cite{Akimov2014}).

A Flexible Surface Hopping scheme was devised in Ref. \cite{Beljonne-SurfaceHopping-JPCLett2013} 
in order to treat large systems, where only a relevant fraction of the
original system is treated with surface hopping with a proper choice of the
time-step. The method was
applied to a model
Hamiltonian similar to that of Eq. (\ref{eq:SSH}) which includes also a coupling
to intramolecular vibrations. 
The resulting mobilities  
show a power law
behavior in the region of interest provided that the coupling with intramolecular
vibrations is not too strong. In the opposite regime of  strong coupling with the 
intramolecular vibrations, instead, the correlation between the lattice distortions and the 
denisty gives rise to small polaron formation,  in which case
a hopping transport regime is achieved. At intermediate coupling strengths a
coexistence of localized and delocalized charges is found with this method, in
agreement with the equilibrium Green's function calculations of Ref. \cite{reconcile09}. 

\paragraph{Lorentzian broadening of the optical conductivity in the static disorder problem.--}
\label{sec:QMC}
The mobility of a system can in principle be obtained from  the   knowledge of the
frequency dependent conductivity. This is given by  the Kubo formula, \cite{Kubo57} which can be stated as
\begin{equation}
\label{eq:Kubo2}
\sigma(\omega) = \frac{1}{\nu \hbar \omega} Re C^R_-(\omega),
\end{equation}
where $C^R_-(\omega)$ is the Fourier transform of the current-current {\it commutator}
correlation function.
The mobility can then be calculated from the
zero-frequency limit of the above expression as \cite{Kubo57}
\begin{equation}
\mu = \lim_{\omega\rightarrow 0^+} \frac{1}{e k_B T \frac{\partial n}{\partial \mu}} 
Re\ \sigma(\omega)
\label{eq:MobilityFiniteDens}
\end{equation}
where $n$ is the carrier density. At low density, the denominator of Eq.
(\ref{eq:MobilityFiniteDens}) reduces to $(e n)$ and one obtains $\sigma_{dc}=ne\mu$.

An  approximation method alternative to the RTA presented above 
has been used in Ref. \cite{KataPRB2011} to evaluate the carrier 
mobility starting from the static disorder 
problem.
Using the Lehman representation for $Re C^R_-(\omega)$ in terms of the
eigenstates $|n\rangle,|m\rangle$ of the {\it static} Hamiltonian ($\omega_0\rightarrow 0$) gives
\begin{eqnarray}
& & Re \ \sigma(\omega)=\frac{2\pi e^2}{\nu \hbar \omega}\sum_{n,m} 
e^{-\beta E_n} \times \label{eq:Leheman} \\ 
& & |\langle n|\hat J|m\rangle |^2 \times (f(E_n)-f(E_m))\delta(\omega-E_m+E_n)
\nonumber
\end{eqnarray}
where   $\hat J$ is current operator and $f(E)$
is the Fermi function. The above expression implies that  $\sigma_{dc}=\sigma(\omega\to 0)=0$, and therefore $\mu=0$.
A finite mobility can be obtained by introducing a  lorentzian broadening
in the delta functions appearing in Eq. (\ref{eq:Leheman}). 

The results obtained by this method are numerically very close 
to the RTA 
(see the data labeled "conv. $\sigma$" in Fig. \ref{fig:mobility}, 
obtained with a broadening $=\hbar\omega_0$). 
This happens because allowing for a finite broadening in the calculation of 
$C^R_-(\omega)$ is mathematically similar to the RTA scheme of 
Sec. \ref{sec:RTA}, where a finite inelastic scattering time was
introduced at the level of the {\it anticommutator} current-current 
correlation function. \cite{PRB12} 
However, the {\it physical content} of the theory is different.
In Ref.
\cite{KataPRB2011} the broadening was assumed to originate from 
the quantum zero-point motion of the molecular vibrations.
This should be contrasted with the RTA equations of Sec. \ref{sec:RTA}, which show that
inelastic scattering processes associated with the molecular fluctuations
are sufficient to provide a finite mobility 
already when the molecular vibrations are classical, $k_BT>\hbar\omega_0$.

\paragraph{QMC with analytical continuation.--}

Recently, a Quantum Monte Carlo (QMC) approach has been applied to the study of the 
model Eq. (\ref{eq:SSH}) by De Filippis {\it  et al.} \cite{KataQMC}
The authors combine diagrammatic \cite{SvistunovDIAGMCPRL1998} 
and worldline Monte Carlo approaches to evaluate the conductivity on the imaginary time axis. 
An analitycal continuation scheme is then used to obtain 
the complex optical conductivity on the real
axis \cite{JarrellMaxEntPhysRep1996} with the aid of exact diagonalization of a small cluster.
This constitutes at present the most complete treatment of quantum effects, and 
it is in principle unbiased. 

The results of this procedure, reported in Fig. \ref{fig:mobility}, fall very close to the RTA, 
Ehrenfest and convolution methods presented above 
(we have rescaled the data of Ref. \cite{KataQMC} 
to the value of the transfer integral $J=143meV$ for comparison
with the other methods). 
Using the
formalism presented in Sec. \ref{sec:optcond}, the authors also calculate the instantaneous
diffusivity, and find a subdiffusive behavior of the system at room temperature, 
in agreement with the transient localization scenario.

We note that 
as was done in Ref. \cite{PRB14} (see next Section), 
Ref. \cite{KataQMC}  proposes a
phenomenological modeling of the optical absorption data.  
De Filippis {\it  et al.}  interpretation of the numerical results is
based on  the assumption that a
large polaron is formed due to the correlation between the carrier  and
the induced molecular deformation, i.e. self-trapping. The finite-frequency absorption would then 
be caused by the internal degrees of freedom of such polaronic particle (which supposedly exists
already at $T=0$),  not by the thermal molecular motions (which instead dominate at room temperature). 
Accordingly, the authors of Ref. \cite{KataQMC}
fit the numerical data based on the Drude-Lorentz model, which highlights the 
existence of an electronic bound  state with a finite radius. 

The whole interpretation in Ref. \cite{KataQMC} therefore relies on the existence of large-polaron 
correlations.
In the considered model Eq. (\ref{eq:SSH})
and at the values of $\lambda$ characteristic of  organic semiconductors, however, the electron is
completely free in the adiabatic regime \cite{CSGSSHPRB1997,DeFilippisPRB2010,KataSSHPRL2010}
and the large-polaron correlations
present at finite values of the phonon frequency are unlikely to survive the
thermal lattice fluctuations at room temperature. A phenomenological model based  on the 
localization induced by thermal molecular motions, and which does not require 
the presence of polaronic electron-lattice correlations,  is presented  in Sec. \ref{sec:phenom} below. \cite{PRB14}



\subsection{Optical conductivity}
\label{sec:optcond}
\subsubsection{Exact relationships} 
An  expression was derived in Ref. \cite{PRB11}
which identifies the quantum spread $\Delta X^2(t)$ in the time domain
as the physical quantity that is dual to the optical conductivity $\sigma(\omega)$ in the frequency domain:
\begin{equation}
   \label{eq:relation}
\sigma(\omega)=-ne^2\omega^2\frac{\tanh 
(\beta\hbar\omega/2)}{\hbar\omega}
   Re \int_0^\infty e^{i\omega t} \Delta X^2(t) dt,
\end{equation}
with $\beta=1/k_BT$ and $n$ the electron density.
This relation, which is a restatement of the Kubo response function theory \cite{Kubo57} based  
on the formal developments by Mayou and collaborators \cite{MayouPRL2000,TramblyPRL2006},
is \textit{exact} for non-degenerate semiconductors. It can be inverted to give
\begin{equation}
   \label{eq:relinv}
  \Delta X^2(t)=-\frac{2 \hbar}{\pi e^2}
  Re \int_0^\infty e^{-i\omega t}\frac{\sigma(\omega)/n}{\omega\tanh (\beta\hbar\omega/2)}
    d\omega.
\end{equation}

This pair of equations shows that the quantum spread of the carriers' wavefunctions 
and the optical response are deeply interconnected
physical quantities. Eq. (\ref{eq:relation})  can be used  to obtain the optical conductivity
$\sigma(\omega)$ from the theoretical knowledge of the time-dependent quantum spread $\Delta X^2(t)$,
i.e. precisely the quantity that is at the core of both the Ehrenfest and the RTA treatments presented above.
For example, via the RTA equations of Sec. \ref{sec:RTA}, one can  easily 
calculate the optical conductivity in the presence of dynamical molecular motions 
from the knowledge of $\sigma(\omega)$
in the limit of static displacements, as was done in Ref. \cite{PRB11}.

Conversely, Eq. (\ref{eq:relinv}) allows to infer the time-dependent quantum dynamics
of charge carriers from the calculated \cite{PRB11,KataQMC} 
or measured \cite{PRB11} optical absorption in an  organic semiconductor. 
Eq. (\ref{eq:relinv}) therefore implies that 
optical absorption measurements are able to
provide fundamental information on the transport mechanism, that are complementary to the 
analysis of the temperature dependence of the mobility \cite{PRB12,PRB14,PRB11}.
In particular, the existence of an intermediate regime of localization between the ballistic 
evolution at short times and the diffusion at long times 
(cf. Fig. \ref{fig:Ehrenfest}) translates, through Eq. (\ref{eq:relinv}),
into a characteristic "Drude-Anderson" optical absorption shape \cite{PRB14}, 
which exhibits a finite frequency peak related to the  transient localization of the carriers  (see next paragraph).
The direct observation of such a finite-frequency peak in the optical absorption measurements performed 
on rubrene single crystals \cite{LiBasov07,FischerAPL2006,YadaAPL2014} (cf. Sec. \ref{sec:exp} and the experimental
data in Fig. \ref{fig:optcond}) therefore provides a strong and direct evidence supporting the transient localization phenomenon.

\subsubsection{A phenomenological model for the analysis of experiments} 
\label{sec:phenom}

\begin{figure}[th!]\center
\includegraphics[width=7cm]{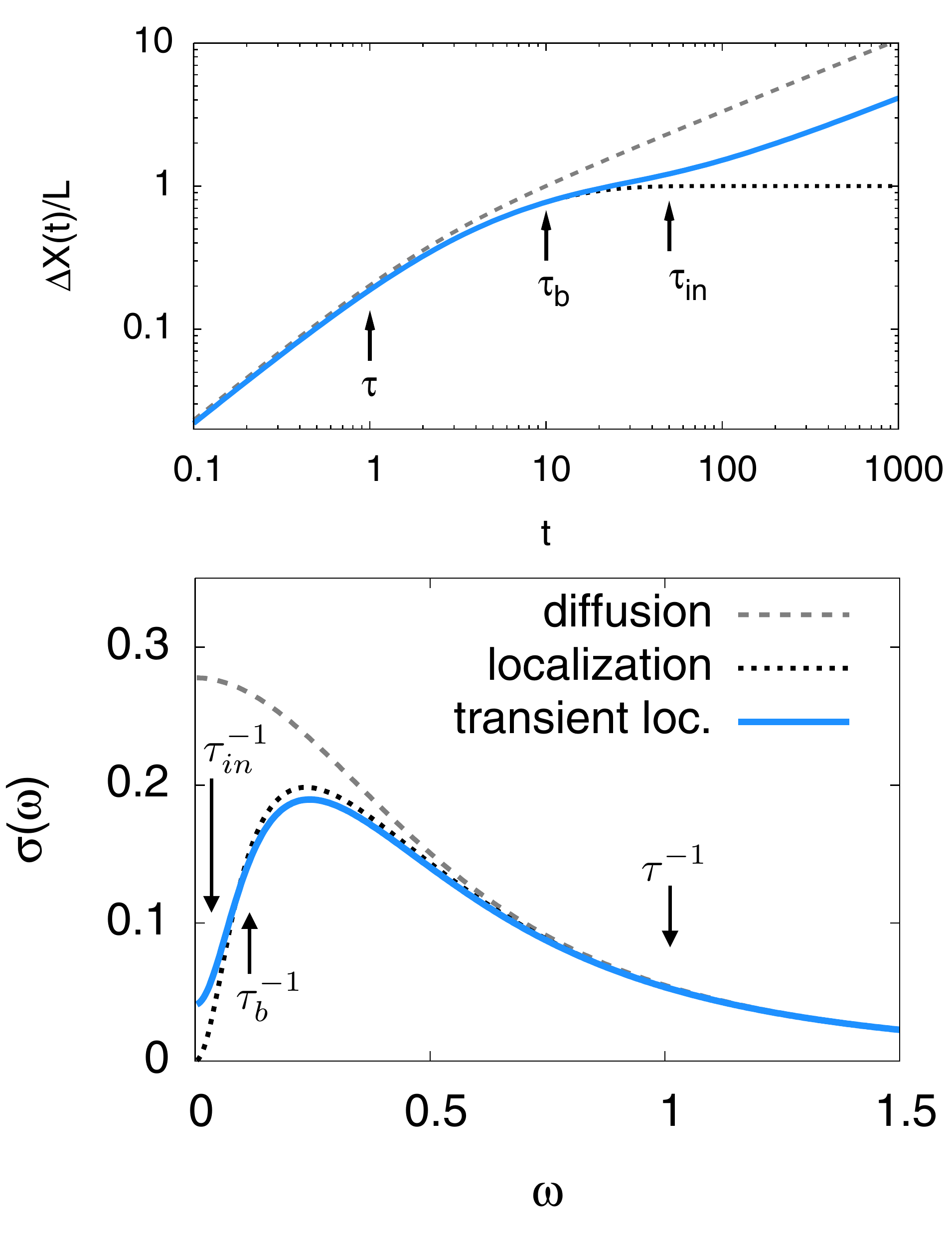}
\caption{
Time dependent quantum spread (top) and optical conductivity (bottom) in the phenomenological
model described by Eqs. (\ref{eq:C0phenom}) and (\ref{eq:resig}). 
The arrows indicate the different timescales of the 
model, here taken as $\tau=1$  (chosen as the time unit; frequencies are 
in units of $1/\tau$), 
$\tau_b=10$, $\tau_{in}=50$, and $k_BT=0.2\hbar/\tau$. 
In both panels, the black dotted line is the localized limit, 
obtained for $\tau_{in}\to\infty$.  The gray dashed line is the Drude-like 
diffusive response alone, corresponding to 
the first term in Eq. (\ref{eq:resig}). Reprinted from Ref. \cite{PRB14}.
}
\label{fig:phenom}
\end{figure}
We have presented in Sec. \ref{sec:RTA} the RTA as a powerful and efficient method
to bridge between the limit of static disorder and the case of dynamical molecular
motions appropriate to  organic semiconductors, and shown that it provides results for the mobility 
that are in good quantitative agreement with the most accurate methods available to date. 
As was pointed out above, however, the greatest success of the RTA scheme is that it provides 
a transparent picture of the charge transport mechanism, showing in simple terms 
how the incipient carrier localization and the dynamics of the molecular motions enter into play. 

The analytical insights developed in Sec. \ref{sec:RTA} can actually be pushed one step forward, 
and used to extract the relevant microscopic parameters
 of charge transport 
directly from the experimental optical absorption data.
To this aim, a phenomenological model was developed in Ref. \cite{PRB14} which provides an analytical
ansatz to the reference correlation function $C_0(t)$ in Eq. (\ref{eq:RTA}), so that 
no numerical calculations are needed throughout the analysis. 
The model is defined in terms of few relevant microscopic parameters already introduced above: 
these are the carrier localization length $L_0$, the elastic scattering time $\tau$, and 
the backscattering time $\tau_b>\tau$ 
encoding the characteristic timescale of the localization process. The correlation function is written as the difference between 
two exponentials, the first representing semi-classical scattering, the second being the 
backscattering term: 
\begin{equation}
C_0(t)=\frac{L_0^2}{\tau_{b}-\tau} \left(\frac{1}{\tau}e^{-t/\tau} -\frac{1}{\tau_{b}}e^{-t/\tau_b}\right).
\label{eq:C0phenom}
\end{equation} 
The  inelastic scattering time $\tau_{in}$ representing the dynamics of disorder
is included via Eq. (\ref{eq:RTA}).
The corresponding time dependent quantum spread is illustrated 
in Fig. \ref{fig:phenom}  (top), and correctly reproduces the qualitative features of
the simulations shown in Fig. \ref{fig:Ehrenfest}.
With this set of parameters, 
 the mobility takes  the RTA form $\mu=(e/k_BT) L^2(\tau_{in})/(2\tau_{in})$ [cf.
Eq. (\ref{eq:muRTA})].
The transient localization length is given by 
$L^2(\tau_{in})=L_0^2/(1+\tau/\tau_{in})(1+\tau_b/\tau_{in})$,
which properly tends to the static localization length $L_0$ when $\tau_{in}\to \infty$.

\begin{figure}[th!]\center
\includegraphics[width=6.cm]{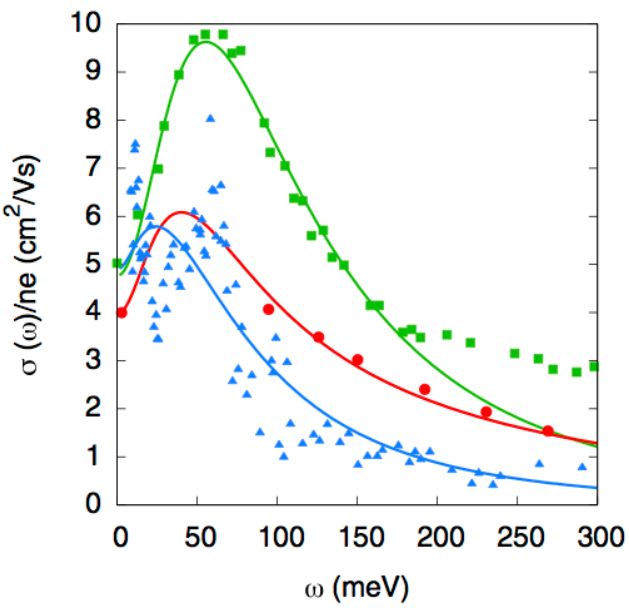}
\includegraphics[width=6.5cm]{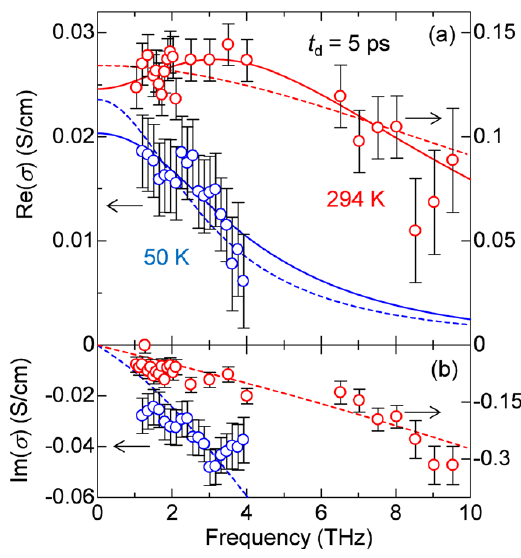}
\caption{Top: Optical conductivity measured in rubrene FETs. 
The full lines are fits through Eq. (\ref{eq:resig}) from Ref. \cite{PRB14}; 
the data are from Refs. \cite{FischerAPL2006} (triangles) \cite{LiBasov07} (squares) and \cite{Okamoto1} (circles). 
Bottom: real and imaginary parts of $\sigma(\omega)$ measured after photoexcitation in Ref. \cite{YadaAPL2014}.
The full line is Eq. (\ref{eq:resig}),
 which properly captures the maximum in the room temperature data, while dashed lines are Drude fits. }
\label{fig:optcond}
\end{figure}

The corresponding optical 
conductivity is obtained in analytical form as: \cite{PRB14}
\begin{eqnarray}
& & Re\ \sigma(\omega)=\frac{ne^2L_0^2}{\tau_b-\tau}\frac{\tanh(\frac{\hbar\omega}{2k_BT})}{\hbar\omega} \times 
\label{eq:resig}\\
& & \times Re \ \left\lbrack
\frac{1}{1+\tau/\tau_{in}-i\omega\tau}-\frac{1}{1+\tau_b/\tau_{in}-i\omega\tau_b}
\right\rbrack. \nonumber 
\end{eqnarray}
Together with the real (dissipative) part, we also report here for the first time 
the  imaginary (refractive) 
part, that is obtained through a Kramers-Kr\"onig transformation. It 
can be expressed via a rapidly converging 
sum over fermionic Matsubara frequencies  $\omega_m=(2m+1)\pi T$ as follows:
\begin{eqnarray}
& & Im\ \sigma(\omega)= \frac{ne^2L_0^2}{\tau_b-\tau} 4k_BT\sum_{m=0}^\infty 
\frac{1}{\omega_m}\frac{\omega}{\omega^2+\omega_m^2} \times \\  \nonumber
& & \times\left\lbrack \frac{1}{1+\tau/\tau_{in}+\omega_m \tau}
-\frac{1}{1+\tau_b/\tau_{in}+\omega_m \tau_b} \right\rbrack. \label{eq:imsig}
\end{eqnarray} 
The above Eqs. (\ref{eq:resig}) and (\ref{eq:imsig})  
can also be generalized to the case of degenerate electron 
systems (see Ref. \cite{PRB14} and Sec. \ref{sec:deg} below).

The present analytical expressions  
have been proven to be  quite accurate to describe 
the region of the Drude-Anderson peak in the optical conductivity of the model Eq. 
(\ref{eq:SSH}), \cite{PRB14} because localization phenomena occur in a frequency range where the 
details of the band dispersion are not important: band effects, that are not contained in the 
analytical ansatz for $C_0$,  only appear at much higher frequencies $\omega\sim J$ 
where the carrier absorption  progressively vanishes anyway. 

Both Eqs. (\ref{eq:resig}) and (\ref{eq:imsig}) 
can be easily implemented in a fitting procedure. 
For example, applying the above Eq. (\ref{eq:resig}) to fit the measurements of Ref. \cite{LiBasov07} nicely reproduces
the shape of the experimental peak (green squares in the top panel of 
Fig. \ref{fig:optcond}), and yields 
the following parameters:  $\hbar/\tau_{in}=13 meV$, $\hbar/\tau_b=40 meV$, 
$\hbar/\tau=195meV$ and $L_0/a=1.9$.
The extracted inelastic scattering rate $\hbar/\tau_{in}$ is 
consistent with the frequency of the intermolecular vibrations in rubrene,
$\omega_0=5-15 meV$ (cf. Sec. \ref{sec:model});
the elastic scattering rate is of the same order of magnitude as the
quasi-particle scattering rates commonly measured in ARPES measurements in  organic semiconductors 
(see e.g. \cite{HatchPRL2010});
the extracted transient localization length indicates that the hole carriers 
are delocalized over few molecules, in agreement with other experimental probes 
 (cf. Sec. \ref{sec:exp}) and with the theoretical results of Fig. \ref{fig:sketch}.

A similar analysis was performed in Ref. \cite{YadaAPL2014} on pump-probe optical conductivity measurements
in rubrene. The analysis of the room temperature data,  shown in the bottom panel of 
Fig. \ref{fig:optcond}, yields
$\hbar/\tau_{in}=8.9 meV$ (again in the correct range of the molecular vibrations), $\hbar/\tau_b=3.4 meV$ and 
$\hbar/\tau=34 meV$. The backscattering rate inferred from the 
position of the peak, $\omega\approx 4THz\sim 12meV$, is much lower than that observed in
rubrene FETs, indicating a reduced level of extrinsic  disorder due to the bulk nature of the pump-probe 
measurement. 
This is also confirmed by the evolution of the spectrum with temperature:
if extrinsic disorder was dominant, the backscattering rate would
increase at low temperatures \cite{PRB14}; instead, the  fit at $T=50K$ yields  
a reduced value $\hbar/\tau_b=2.7 meV$, demonstrating that the pump-probe measurements are 
free from interface effects and are actually 
probing the intrinsic carrier dynamics in the  organic semiconductors.




\subsection{Degenerate systems}
\label{sec:deg}
Although in this overview we have mainly focused on the transient localization phenomenon in organic semiconductors, 
the concept itself is much more general and applies as well to other classes of materials.

 The interplay between Anderson localization and lattice vibrations was  studied in the
past in the different framework of random metal alloys
and other degenerate disordered systems. 
It was recognized early on by  Gogolin and collaborators \cite{GogolinJETP75,RashbaGogolinMelnikov76} and Thouless
\cite{ThoulessPRL1977} that the random fluctuations introduced by the lattice motions 
destroy the quantum interferences necessary for localization of the electronic states, which is precisely 
the idea underlying the RTA treatment presented in Sec. \ref{sec:RTA}.
In the scaling theories of
localization, \cite{LeeRMP1985} the inelastic scattering by dynamical lattice motions is   included as a
cutoff for localization corrections, allowing an otherwise localized electron system 
to support a finite electrical conductivity. Beyond such scaling arguments, 
a microscopic calculation of the effects of lattice dynamics on charge transport 
in strongly disordered systems based on the Kubo formula was provided by Girvin and Jonson. \cite{GirvinPRL1979}

 The case of one-dimensional disordered 
systems was studied in depth by several authors via diagrammatic techniques 
\cite{RashbaGogolinMelnikov76,MadhukarCohenPRL1977, GogolinMelnikovRashbaJETP77,GogolinPhysRep81}.
These authors provided an estimate for the diffusivity  equivalent to Eq. (\ref{eq:DRTA}),
\begin{equation} 
D\sim \frac{L^2}{\tau_{in}}.
\end{equation} 
A physical interpretation of this formula was given in terms 
of electrons being localized at intermediate time-scales by the dynamic disorder 
introduced by the
phonons, which is in essence equivalent to the transient localization mechanism discussed in this review.
The concept was also generalized to 2D systems in Refs. \cite{GogolinZimanyiSSC83,GogolinZimanyiSSC84}.

Interestingly, these ideas were actually applied  \cite{RashbaGogolinMelnikov76,ShanteJPC1978,GogolinPhysRep81}
to the analysis of the transport and optical properties of 
both one- and two-dimensional organic conductors, taking the compound TTF-TCNQ as a paradigmatic case. 
Low-dimensional organic conductors can be viewed as the doped (degenerate) analogues of the  organic semiconductors discussed in this overview.
In particular, they have narrow bands constructed from the pi-overlaps between adjacent organic molecules,
and therefore the effects of inter-molecular motions should be analogous to those in  organic semiconductors.
Although in such degenerate electron systems
collective effects related to electron correlations can enter into play at low temperatures 
\cite{ShanteJPC1978,TakenakaPRL05,SeoJPSJ2006,CanoPRL2010}, the effects of thermal lattice motions 
should become dominant at high temperatures, and
it is not suprising that also in this class of materials several distinctive 
features  of the transient localization mechanism
are commonly observed: metallic-like power-law temperature dependence of the  conductivity, 
with low values $\sigma = 10-1000 (\Omega cm)^{-1}$ \cite{Graja} corresponding to room temperature
mobilities in the range  $\mu = 0.1-10cm^2/Vs$;  
breakdown of the MIR condition; apparent localization lengths of few molecular units; and
non-Drude-like optical conductivities exhibiting  marked finite-frequency peaks in the infra-red region
\cite{TakenakaPRL05,Hashimoto}
(see \cite{PRB14} for a detailed discussion).

It would be interesting to apply the  full phenomenological model of Sec. \ref{sec:phenom} to 
perform a systematic analysis of the experimental results in low-dimensional organic conductors \cite{Dressel}
and other classes of compounds. 
To this aim we report here the formulas 
which generalize Eqs. (\ref{eq:resig}) and (\ref{eq:imsig})   to
degenerate electron systems in the low temperature limit: \cite{PRB14}
\begin{eqnarray}
& & \sigma(E_F,\omega)=e^2N(E_F)\frac{C(E_F,0)}{1/\tau-1/\tau_b} \times
\label{eq:transloc2}\\
& & \times  \left\lbrack
\frac{1}{1+\tau/\tau_{in}-i\omega\tau}-\frac{1}{1+\tau_b/\tau_{in}-i\omega\tau_b}
\right\rbrack \nonumber 
\end{eqnarray}
with $N(E_F)$ the density of states at the Fermi energy and $C(E_F,0)$ the short-time limit of the 
velocity correlation function of electrons at the Fermi energy. We conclude this section with some remarks.

(i) The above formula has been used in Ref. \cite{PRB14} to fit the room temperature 
optical conductivity data in 
the two-dimensional compound $\theta$-ET$_2$I$_3$ \cite{TakenakaPRL05} (a compound where 
$\sigma = 10 (\Omega cm)^{-1}$ at room temperature,  below the Mott-Ioffe-Regel limit \cite{GunnarssonNat2000}). 
A backscattering rate $\hbar/\tau_b\simeq 14 meV$  could be extracted, as well as
an elastic scattering rate $\hbar/\tau=116 meV$, both in the correct range
expected from electron-molecular vibration coupling. An analogous  absorption peak 
in the far infrared range has also been observed in the compound $\theta$-ET$_2$CsZn, which presents a glassy
electronic state. \cite{Hashimoto}  In that case, in addition to the dynamical molecular disorder, the 
effect of the random electrostatic potentials of the electrons in the glassy configurations \cite{Mahmoudian}
could also give rise to an absorption shape of the form of Eq. (\ref{eq:transloc2}).

(ii) Ideally, the realization of FETs with ionic liquid gating can 
bridge continuously  from the physics of non-degenerate  organic semiconductors to that of degenerate
organic conductors. It has been shown recently in Ref. \cite{XiePRL2014}
that in rubrene, carrier densities up to $6\times 10^{13} cm^{-2}$ can be reached  by this technique 
(0.3 holes per molecule), where not only the Fermi-Dirac statistics but also 
many-body electron correlation effects become important. It would be extremely interesting to track 
the evolution of the electronic properties
as a function of carrier concentration in such devices.

\begin{figure}[th!]\center
\includegraphics[width=6.5cm]{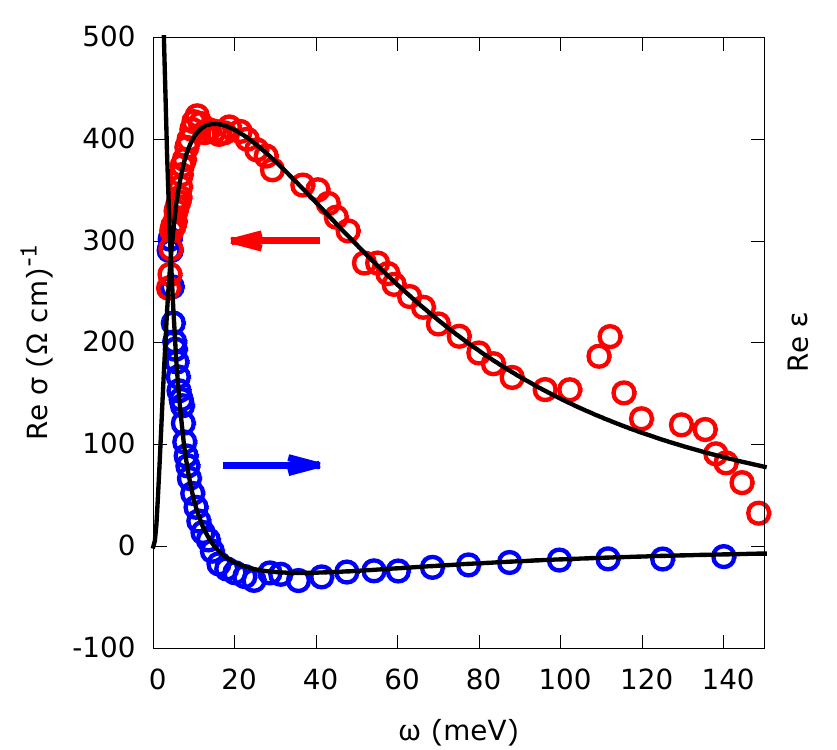}
\caption{Real part of the optical conductivity $\textrm{Re} \sigma$ 
and dielectric constant $Re \ \varepsilon=1-4\pi \textrm{Im} \sigma/\omega$ (in arb. units)
measured in carbon nanotubes, 
simultaneously fitted through Eq. (\ref{eq:transloc2}). Data are from Ref. \cite{MartelPRL2008}. }
\label{fig:nanotubes}
\end{figure}

(iii) In a different class of low-dimensional materials ---
carbon nanotubes --- an ubiquitous optical absorption peak  in the far infrared range
has been reported by several groups 
\cite{MartelPRL2008,IchidaSSC2011,UlbrichtRMP2011}, whose microscopic origin could possibly
be related to the transient localization phenomena described in this work. Fig.
\ref{fig:nanotubes} illustrates the (simultaneous) fits of both the real and
imaginary part of the optical conductivity measured in Ref. \cite{MartelPRL2008} via Eq. (\ref{eq:transloc2}),
showing an excellent agreement with the data. The extracted values are
$\hbar/\tau_b\simeq 3.3 meV$ and $\hbar/\tau=68 meV$ ($\tau_{in}\gg \tau_b,\tau$
was assumed).


\section{Outlook}
\label{sec:out}


Understanding the intrinsic charge transport
mechanism in high-mobility organic semiconductors requires a theory that is able to reconcile the  
apparent contradiction between the "band-like" temperature dependence of
the mobility, suggestive of the existence of extended charge carriers, 
and the presence of localization phenomena as seen in a variety of experiments.
It is now ascertained that such duality originates, at the microscopic level, from the 
presence of large thermal molecular motions. 
Dynamical deviations from the perfect crystalline arrangement 
act as a strong source of disorder,  inducing  a
localization of the electronic wave functions on the typical timescales of the
inter-molecular vibrations. Such  \textit{transient localization} is what 
limits the room temperature mobilities  down to few tens of $cm^2/Vs$, as observed in the best 
organic semiconductors. 


In this article, we have provided an overview of the
different theoretical approaches that have been applied to the problem, 
focusing on an extensively studied model, Eq. (\ref{eq:SSH}),
 which describes the interaction of the charge 
carriers with the inter-molecular motions. 
The standard treatments  based either on 
semi-classical band transport or polaron hopping,  presented in Sec. \ref{sec:early},  
have been shown to be unable to provide a satisfactory 
description of the experiments, because they cannot capture the
quantum localization effects caused by the large molecular motions.
The corresponding results for the temperature dependence of the mobility 
are summarized in Fig. \ref{fig:semic}: while a band-like power-law dependence is commonly obtained, 
these methods generally overestimate the absolute value of the mobility, in the worst cases by 
a full order of magnitude.

Sec. \ref{sec:transloc} describes a number of  theoretical approaches that have been 
developed recently and that can  properly account for 
the effects of strong dynamical disorder. 
All these methods, which give direct access
to the time evolution of the electronic wavefunction,
indicate that the carriers become localized up to timescales corresponding to 
the period of the molecular oscillations.
Such transient localization results  in an original regime of charge transport 
where the  carriers exhibit  both localized and extended characters, as observed in experiments.

Sec. \ref{sec:transloc} also contains a description of some important analytical developments which
help shed light on the microscopic transport mechanism at work in  organic semiconductors.
First, by taking quantum localization effects as a starting point,
the relaxation time approximation described in Sec. \ref{sec:RTA}
allows to understand in simple physical terms how the transient localization
caused by dynamical molecular motions relates to the
Anderson localization realized for static disorder.
Next, the general theoretical framework presented in Sec. \ref{sec:optcond}  identifies 
the optical conductivity in the frequency domain as 
a physical quantity that provides crucial information on the 
charge transport mechanism,  complementary to the temperature dependence of the mobility. 
In particular, it is shown that the transient localization phenomenon is directly reflected
in the emergence of a finite-frequency peak in the optical conductivity, 
whose existence in  organic semiconductors has been confirmed experimentally by different groups.

In the regime of microscopic parameters appropriate to high mobility organic semiconductors, all the
modern  theoretical approaches presented in Sec. \ref{sec:transloc}
yield quantitatively comparable results 
for the temperature dependence and absolute value of the charge carrier mobility.
As shown in Fig. \ref{fig:mobility}, a remarkable agreement 
is found with the experimental data available in rubrene FETs,
especially considering the simplicity of the model Eq. (\ref{eq:SSH}).
%
With the limitations described in Sec. \ref{sec:overview} in mind,
the calculations presented in Sec. \ref{sec:transloc} 
thus have a real predictive power. Moreover,
having identified  how the different microscopic parameters enter in the 
charge transport processes allows one to devise efficient strategies to improve the performances
of actual organic  devices.
According to Fig. \ref{fig:mobility}, for example, 
if all sources of extrinsic disorder were removed 
one could attain mobilities above $100 cm^2/Vs$ upon lowering the temperature below $100K$. 
From the analytical arguments given in  Sec. \ref{sec:RTA}, seeking compounds with 
tighter inter-molecular bonds in order to reduce the inter-molecular fluctuations 
  appears as a promising route
to improve the transport characteristics of  organic semiconductors.

Finally, it appears the general phenomenon of transient localization, discussed here in the framework of
the crystalline organic semiconductors, is actually relevant in
broader classes of materials such as low-dimensional organic metals, as well as 
glassy and disordered systems. A similar
concept of environment-assisted quantum transport has also been put forward to explain charge transfer 
in biological light-harvesting systems \cite{Engel-Nature07,Mohseni-Rebentrost-JCP08,Rebentrost-NJP09}.


\bigskip
\hrulefill
\bibliographystyle{ciuk}

\footnotesize
\begin{thebibliography}{100}

\bibitem{BredasChemRev04}
J.-L. Br\'edas, D. Beljonne, V. Coropceanu,  J. Cornil,
\newblock {\em Chem. Rev.}, {\bf 2004}, 104, 4971.


\bibitem{BredasACR2009}
J.-L. Br\'edas, J. E. Norton, J. Cornil,  V. Coropceanu,
\newblock {\em Accounts Chem. Res.}, {\bf 2009}, 42, 1691.

\bibitem{MinderAdvMat2011}
N.~A. Minder, S. Ono, Z. Chen, A. Facchetti, A.~F.
  Morpurgo,
\newblock {\em Adv. Mater.}, {\bf 2012}, 24, 503.

\bibitem{SirringhausNatMat2010}
T. Sakanoue, H. Sirringhaus,
\newblock {\em Nat. Mater}, {\bf 2010}, 9, 736.

\bibitem{DukeSchein80}
C.~B. Duke, L.~B. Schein.
\newblock {\em Phys. Today}, {\bf 1980}, 33, 42.

\bibitem{DevosPRB1998}
A.~Devos, M.~Lannoo.
\newblock {\em Phys. Rev. B}, {\bf 1998}, 58, 8236.

\bibitem{CoropceanuPRL2002}
V.~Coropceanu, M.~Malagoli, D.~A. da~Silva~Filho, N.~E. Gruhn, T.~G. Bill, 
  J.~L. Br\'edas.
\newblock {\em Phys. Rev. Lett.}, {\bf 2002}, 89, 275503.

\bibitem{BlaseEphPRB2011}
C. Faber, J. Laflamme Janssen, M. C\^ot\'e, E.~Runge, 
  X.~Blase.
\newblock {\em Phys. Rev. B}, {\bf  2011}, 84, 155104.

\bibitem{GirlandoPRB10}
A. Girlando, L. Grisanti, M. Masino, I. Bilotti, A. Brillante,
  R.~G. Della~Valle,  E. Venuti.
\newblock {\em Phys. Rev. B}, {\bf  2010}, 82, 035208.

\bibitem{SinovaPRL2001}
J. Sinova, J. Schliemann, A.~S. N\'u\~nez,  A.~H. MacDonald,
\newblock {\em Phys. Rev. Lett.}, {\bf  2001}, 87, 226802.

\bibitem{TroisiPRL06}
A. Troisi, G. Orlandi,
\newblock {\em Phys. Rev. Lett.}, {\bf 2006}, 96, 086601.

\bibitem{LeeRMP1985}
P.~A. Lee, T.~V. Ramakrishnan.
\newblock {\em Rev. Mod. Phys.}, {\bf  1985}, 57, 287.

\bibitem{BasslerReviewPSS1993}
H.~B\"assler,
\newblock {\em phys. status solidi (b)}, {\bf 1993}, 175, 15.

\bibitem{CoehoornPRB2005}
R.~Coehoorn, W.~F. Pasveer, P.~A. Bobbert, M.~A.~J. Michels.
\newblock {\em Phys. Rev. B}, {\bf  2005}, 72, 155206.

\bibitem{FishchukPRB2009}
I.I.~Fishchuk, V.~I.~Arkhipov, A.~Kadashchuk, P.~Heremans, H.~B{\"a}ssler 
\newblock {\em Phys. Rev. B}, {\bf 2007}, 76, 045210.

\bibitem{Glarum63}
S.~H. Glarum,
\newblock {\em J. Phys. Chem. Solids}, {\bf 1963},
  24, 1577.

\bibitem{FriedmanPR65}
L. Friedman,
\newblock {\em Phys. Rev.}, {\bf  1965}, 140, A1649.

\bibitem{GosarChoiPR1966}
P.~Gosar and Sang-il Choi.
\newblock {\em Phys. Rev.}, {\bf Oct 1966}, 150:529--538.

\bibitem{MadhukarPostPRL1977}
A.~Madhukar and W.~Post.
\newblock {\em Phys. Rev. Lett.}, {\bf Nov 1977}, 39:1424--1427.

\bibitem{Laarhoven-JCP08}
H.~A. v.~Laarhoven, C.~F.~J. Flipse, M.~Koeberg, M.~Bonn, E.~Hendry,
  G.~Orlandi, O.~D. Jurchescu, T.~T.~M. Palstra,  A.~Troisi,
\newblock {\em J. of Chem. Phys.}, {\bf 2008}, 129, 044704.

\bibitem{SirringhausPSS2012}
H. Sirringhaus, T. Sakanoue,  J.-F. Chang,
\newblock {\em phys. status solidi (b)}, {\bf 2012}, 249, 1655.

\bibitem{Chang-Troisi-SirringhausPRL2011}
J.-F. Chang, T. Sakanoue, Y. Olivier, T. Uemura, M.-B.
  Dufourg-Madec, S.~G. Yeates, J. Cornil, J. Takeya, A.
  Troisi, H. Sirringhaus,
\newblock {\em Phys. Rev. Lett.}, {\bf  2011}, 107, 066601.

\bibitem{SirringhausNatMat2013}
A.~S. Eggeman, S. Illig, A. Troisi, H. Sirringhaus,
   P.~A. Midgley,
\newblock {\em Nat. Mater.}, {\bf  2013}, 12, 1045.

\bibitem{YadaAPL2014}
H.~Yada, R.~Uchida, H.~Sekine, T.~Terashige, S.~Tao, Y.~Matsui, N.~Kida,
  S.~Fratini, S.~Ciuchi, Y.~Okada, T.~Uemura, J.~Takeya, and H.~Okamoto,
\newblock {\em Appl. Phys. Lett.}, {\bf 2014}, 105, 143302.

\bibitem{MinderAdvMat2014}
N.~A. Minder, S. Lu, S. Fratini, S. Ciuchi, A.
  Facchetti,  A.~F. Morpurgo,
\newblock {\em Adv. Mater.}, {\bf 2014}, 26, 1254.

\bibitem{TroisiPCCP08}
D.~L. Cheung, A. Troisi,
\newblock {\em Phys. Chem. Chem. Phys.}, {\bf 2008}, 10, 5941.

\bibitem{TroisiChemSocRev2011}
A. Troisi,
\newblock {\em Chem. Soc. Rev.}, {\bf 2011}, 40, 2347.

\bibitem{CoropceanuChemRev2007}
V. Coropceanu, J. Cornil, D.~A. da~Silva~Filho, Y. Olivier,
  R. Silbey, J.-L. Br\'edas,
\newblock {\em Chem. Rev.}, {\bf 2007}, 107, 926.


\bibitem{Schweicher-IJCH2014}
G. Schweicher, Y. Olivier, V. Lemaur,  Y.~H. Geerts,
\newblock {\em Israel J. Chem.}, {\bf 2014}, 54, 595.

\bibitem{OrgElII}
\newblock {\em Organic Electronics II} (Ed. Hagen Klauk), 
\newblock Wiley-VCH Verlag Gmbh \& Co. KGaA, {\bf 2012}.

\bibitem{Shuai-review-AdvMat11}
Z. Shuai, L. Wang,  Q. Li,
\newblock {\em Adv. Mater.}, {\bf 2011}, 23, 1145.

\bibitem{StafstromChemSocRev2010}
S. Stafstrom,
\newblock {\em Chem. Soc. Rev.}, {\bf 2010}, 39, 2484.

\bibitem{PernstichNMat2008}
K.~P. Pernstich, B.~Rossner,  B.~Batlogg.
\newblock {\em Nat. Mater.}, {\bf  2008}, 7, 321.

\bibitem{PodzorovHallPRL2005}
V.~Podzorov, E.~Menard, J.A.~Rogers,  M.E.~Gershenson,
\newblock {\em Phys. Rev. Lett.}, {\bf  2005}, {95}, 226601.

\bibitem{KalbPRB2010}
W.~L. Kalb, S. Haas, C. Krellner, T. Mathis,  B.
  Batlogg,
\newblock {\em Phys. Rev. B}, {\bf  2010}, 81, 155315.

\bibitem{WillaJAP2013}
K.~Willa, R.~H\"ausermann, T.~Mathis, A.~Facchetti, Z.~Chen,  B.~Batlogg,
\newblock {\em J. Appl. Phys.}, {\bf 2013}, 113, 133707.

\bibitem{FischerAPL2006}
M. Fischer, M. Dressel, B. Gompf, A.~K. Tripathi,  J.
  Pflaum,
\newblock {\em Appl. Phys. Lett.}, {\bf 2006}, 89, 182103.

\bibitem{LiBasov07}
Z.~Q. Li, V.~Podzorov, N.~Sai, M.~C. Martin, M.~E. Gershenson, M.~Di Ventra,
   D.~N. Basov,
\newblock {\em Phys. Rev. Lett.}, {\bf 2007}, 99, 016403.

\bibitem{MarumotoPRL2006}
K. Marumoto, S.-I.  Kuroda, T. Takenobu,  Y. Iwasa,
\newblock {\em Phys. Rev. Lett.}, {\bf  2006}, 97, 256603.

\bibitem{MatsuiPRL2010}
H. Matsui, A.~S. Mishchenko,  T. Hasegawa,
\newblock {\em Phys. Rev. Lett.}, {\bf  2010}, 104, 056602.

\bibitem{Marumoto11}
K. Marumoto, N. Arai, H. Goto, M. Kijima, K.
  Murakami, Y. Tominari, J. Takeya, Y. Shimoi, H. Tanaka,
  S.-I. Kuroda, T. Kaji, T. Nishikawa, T. Takenobu, 
  Y. Iwasa,
\newblock {\em Phys. Rev. B}, {\bf  2011}, 83, 075302.

\bibitem{Holstein-1959}
a) T.~Holstein,
\newblock {\em Annals of Physics}, {\bf 1959}, 8, 325; b)
T.~Holstein,
\newblock {\em Annals of Physics}, {\bf 1959}, 8, 343.

\bibitem{PiovraEPL1998}
M.~{Capone}, S.~{Ciuchi},  C.~{Grimaldi},
\newblock {\em Europhys. Lett.}, {\bf  1998}, 42, 523.

\bibitem{rhopolaron}
S.~{Fratini}, S.~{Ciuchi},
\newblock {\em Phys. Rev. Lett.}, {\bf  2003}, 91, 256403.

\bibitem{Beljonne-SurfaceHopping-JPCLett2013}
L. Wang,  D. Beljonne,
\newblock {\em J. Phys. Chem. Lett.}, {\bf 2013},
  4, 1888.

\bibitem{KataHolstein2015}
S.~Mishchenko, A.\, N.~Nagaosa, G.~De~Filippis, A.~de~Candia, 
  V.~Cataudella,
\newblock {\em Phys. Rev. Lett.}, {\bf  2015}, 114, 146401.

\bibitem{VukmirovicPRL2012}
N Vukmirovi\'{c}, C.~Bruder, V.~M.
  Stojanovi\'{c},
\newblock {\em Phys. Rev. Lett.}, {\bf  2012}, 109, 126407.

\bibitem{orgarpes}
S.~Ciuchi, S.~Fratini,
\newblock {\em Phys. Rev. Lett.}, {\bf  2011}, 106, 166403.

\bibitem{exporgarpes}
S.~Ciuchi, R.~C. Hatch, H.~H\"ochst, C.~Faber, X.~Blase,  S.~Fratini,
\newblock {\em Phys. Rev. Lett.}, {\bf  2012}, 108, 256401.

\bibitem{ZuppiroliEPL2004}
M.~N. Bussac, J.~D. Picon,  L.~Zuppiroli,
\newblock {\em Europhys. Lett.}, {\bf  2004}, 66, 392.

\bibitem{PiconPRB2007}
J.-D. Picon, M.~N. Bussac,  L.~Zuppiroli,
\newblock {\em Phys. Rev. B}, {\bf 2007}, 75, 235106.

\bibitem{PRB12}
S.~Ciuchi, S.~Fratini.
\newblock {\em Phys. Rev. B}, {\bf  2012}, 86, 245201.

\bibitem{KirovaPRB2003}
N.~Kirova, M.-N. Bussac.
\newblock {\em Phys. Rev. B}, {\bf  2003}, 68, 235312.

\bibitem{NatMat}
I.~N. {Hulea}, S.~{Fratini}, H.~{Xie}, C.~L. {Mulder}, N.~N. {Iossad},
  G.~{Rastelli}, S.~{Ciuchi}, A.~F. {Morpurgo}, 
\newblock {\em Nat. Mater.}, {\bf  2006}, 5, 982.

\bibitem{RichardsJCP2008}
T.  Richards, M.  Bird, H. Sirringhaus, 
\newblock {\em J. of Chem. Phys.}, {\bf 2008}, 128, 234905.

\bibitem{DaSilvaAdvMat2005}
D.~A. da~Silva~Filho, E.-G. Kim,  J.-L. Bredas,
\newblock {\em Adv. Mater.}, {\bf 2005}, 17, 1072.

\bibitem{TroisiAdvMat2007}
A.~Troisi, 
\newblock {\em Adv. Mater.}, {\bf 2007}, 19, 2000.

\bibitem{HannewaldBobbertAPL2004}
K.~Hannewald, P.~A. Bobbert,
\newblock {\em Appl. Phys. Lett.}, {\bf 2004}, 85, 1535.

\bibitem{HannewaldPRB2004}
K.~Hannewald, Stojanovi\'{c}~V. M., J.~M.~T. Schellekens, P.~A. Bobbert,
  G.~Kresse,  J.~Hafner,
\newblock {\em Phys. Rev. B}, {\bf  2004}, 69, 075211.

\bibitem{KatoJCP2002}
T.  Kato, K.  Yoshizawa,  K. Hirao,
\newblock {\em J. Chem. Phys.}, {\bf 2002}, 116, 3420.

\bibitem{SanchezCarrera-JACS10}
R. ~S. S\'anchez-Carrera, P.  Paramonov, G. ~M. Day, V.  Coropceanu,
  J.-L.  Br\'edas,
\newblock {\em J. Am. Chem. Soc.}, {\bf 2010},
  132, 14437.

\bibitem{reconcile09}
S.~Fratini, S.~Ciuchi.
\newblock {\em Phys. Rev. Lett.}, {\bf  2009}, 103, 266601.

\bibitem{SSH+KivelsonRMP1088}
A.~J. Heeger, S.~Kivelson, J.~R. Schrieffer,  W.~P. Su,
\newblock {\em Rev. Mod. Phys.}, {\bf  1988}, 60, 781.

\bibitem{CSGSSHPRB1997}
M.~Capone, W.~Stephan,  M.~Grilli,
\newblock {\em Phys. Rev. B}, {\bf  1997}, 56, 4484.

\bibitem{ZoliSSHacuPertPRB2002}
M. Zoli,
\newblock {\em Phys. Rev. B}, {\bf  2002}, 66, 012303.

\bibitem{SakaiPRB12}
Y.~Okada, K.~Sakai, T.~Uemura, Y.~Nakazawa,  J.~Takeya,
\newblock {\em Phys. Rev. B}, {\bf  2011}, 84, 245308.

\bibitem{PodzorovPRL2004}
V.~Podzorov, E.~Menard, A.~Borissov, V.~Kiryukhin, J.A.~Rogers,  M.E.~Gershenson,
\newblock {\em Phys. Rev. Lett.}, {\bf  2004}, 93, 086602.

\bibitem{FrisbieJPC2013}
W.  Xie, K.~A. McGarry, F. Liu, Y. Wu, P.~P. Ruden,
  C.~J. Douglas,  C.~D. Frisbie,
\newblock {\em J. Phys. Chem. C}, {\bf 2013},
  117, 11522.

\bibitem{Mahan}
G.~D. Mahan,
\newblock {\em Many-Particle Physics}.
\newblock Kluwer Academic/Plenum Publisher, New Yok, {\bf 2000}.

\bibitem{PRB14}
S.~Fratini, S.~Ciuchi,  D.~Mayou,
\newblock {\em Phys. Rev. B}, {\bf  2014}, 89, 235201.

\bibitem{ChengJCP2003}
Y.~C. Cheng, R.~J. Silbey, D.~A. da~Silva~Filho, J.~P. Calbert, J.~Cornil, 
  J.~L. Br\'{e}das,
\newblock {\em J. Chem. Phys.}, {\bf 2003}, 118, 3764.

\bibitem{MillisPRL1999}
A.~J. Millis, J. Hu,  S.~Das~Sarma, 
\newblock {\em Phys. Rev. Lett.}, {\bf  1999}, 82, 2354.

\bibitem{GunnarssonNat2000}
a) O.~Gunnarsson, J.~E. Han,
\newblock {\em Nature}, {\bf  2000}, 405, 1027;
b) O.~Gunnarsson, M.~Calandra,  J.~E. Han,
\newblock {\em Rev. Mod. Phys.}, {\bf  2003}, 75, 1085.

\bibitem{SumiSSHJCP79}
H.~Sumi, 
\newblock {\em J. Chem. Phys.}, {\bf 1979}, 70, 3775.

\bibitem{Kubo57}
R.~Kubo,
\newblock {\em J. Phys. Soc. Japan.}, {\bf 1957}, 12, 570.

\bibitem{TroisiJCP2011}
A. Troisi,
\newblock {\em J. Chem. Phys.}, {\bf 2011}, 134, 034702.

\bibitem{AustinMott}
I.G. Austin,  N.F. Mott, 
\newblock {\em Adv. Phys.}, {\bf 1969}, 18, 41.

\bibitem{MunnSilbeyJCP1985}
R.~W. Munn,  R.~Silbey,
\newblock {\em J. Chem. Phys.}, {\bf 1985}, 83, 1843.

\bibitem{KenkrePRL89}
V.~M. Kenkre, J.~D. Andersen, D.~H. Dunlap,  C.~B. Duke,
\newblock {\em Phys. Rev. Lett.}, {\bf  1989}, 62, 1165.

\bibitem{OrtmannAPL2008}
F. Ortmann, K. Hannewald,  F. Bechstedt,
\newblock {\em Appl. Phys. Lett.}, {\bf 2008}, 93, 222105.

\bibitem{OrtmannPRB2009}
F. Ortmann, F. Bechstedt,  K. Hannewald,
\newblock {\em Phys. Rev. B}, {\bf  2009}, 79, 235206.

\bibitem{LangFirsovZETP1962}
I.~G. Lang, Yu.~A. Firsov.
\newblock {\em Zh. Eksp. Teor. Fiz.}, {\bf 1962}, 43, 1843.

\bibitem{LangFirsovJETP1963}
I.~G. Lang, Yu.~A. Firsov.
\newblock {\em Sov. Phys. JETP}, {\bf 1963}, 16, 1301.


\bibitem{HannewaldStojanovicJPC2004}
K.~Hannewald, V.~M. Stojanov\'{c},  P.~A. Bobbert,
\newblock {\em J. Phys.-Condens. Mat.}, {\bf 2004}, 16, 2023.

\bibitem{RanningerPRB1993}
J. Ranninger,
\newblock {\em Phys. Rev. B}, {\bf  1993}, 48, 13166.

\bibitem{AlexKornilovicPRL1999}
A.~S. Alexandrov,  P.~E. Kornilovitch,
\newblock {\em Phys. Rev. Lett.}, {\bf  1999}, 82, 807.

\bibitem{BerciuPRL2006}
M. Berciu,
\newblock {\em Phys. Rev. Lett.}, {\bf 2006}, 97, 036402.

\bibitem{HatchPRL2010}
R.~C. Hatch, D.~L. Huber,  H.t H\"ochst,
\newblock {\em Phys. Rev. Lett.}, {\bf  2010}, 104, 047601.

\bibitem{MasinoMS2004}
M. Masino, A. Girlando, A. Brillante, L. Farina, R.~G.
  Della~Valle,  E. Venuti,
\newblock {\em Macromol. Sy.}, {\bf 2004}, 212, 375.

\bibitem{LiNarrowingJPCL2012}
Y. Li, V. Coropceanu, J.-L. Br\'edas,
\newblock {\em J. of Phys. Chem. Lett.}, {\bf 2012},
  3, 3325.

\bibitem{Li-Coropceanu-PRB12}
Y. Li, Y. Yi, V. Coropceanu, J.-L. Br\'edas,
\newblock {\em Phys. Rev. B}, {\bf  2012}, 85, 245201.

\bibitem{PRB11}
S.~Ciuchi, S.~Fratini,  D.~Mayou,
\newblock {\em Phys. Rev. B}, {\bf  2011}, 83, 081202.

\bibitem{MozafariStafstromJCP2013}
E.~Mozafari,  S.~Stafstrom, 
\newblock {\em J. Chem. Phys.}, {\bf 2013}, 138, 184104.

\bibitem{DeFilippisPRB2010}
G.~De~Filippis, V.~Cataudella, S.~Fratini,  S.~Ciuchi,
\newblock {\em Phys. Rev. B}, {\bf  2010}, 82, 205306.

\bibitem{WangPCCP2010}
L. Wang, Q. Li, Z. Shuai, L. Chen,  Q. Shi,
\newblock {\em Phys. Chem. Chem. Phys.}, {\bf 2010}, 12, 3309.

\bibitem{Wang-BeljonneJCP2011}
L. Wang, D. Beljonne, L. Chen, Q. Shi,
\newblock {\em J. Chem. Phys.}, {\bf 2011}, 134, 244116.

\bibitem{Ishii-MobilityWavePacket-PRB2012}
H. Ishii, K. Honma, N. Kobayashi,  K. Hirose,
\newblock {\em Phys. Rev. B}, {\bf  2012}, 85, 245206.

\bibitem{Tamura-Eherenfest2d-PRB2012}
H. Tamura, M. Tsukada, H. Ishii, N. Kobayashi,  K.
  Hirose,
\newblock {\em Phys. Rev. B}, {\bf  2012}, 86, 035208.

\bibitem{Ishii-Ehrenfest-2D-anisotropy-PRB13}
H. Ishii, N. Kobayashi, K. Hirose,
\newblock {\em Phys. Rev. B}, {\bf  2013}, 88, 205208.

\bibitem{Ishii-Hall-PRB2014}
H.  Ishii, H. Tamura, M. Tsukada, N. Kobayashi, K.
  Hirose,
\newblock {\em Phys. Rev. B}, {\bf  2014}, 90, 155458.

\bibitem{ParandekarJCP2005}
P.~V. Parandekar,  J.~C. Tully,
\newblock {\em J. Chem. Phys.}, {\bf 2005}, 122, 094102.

\bibitem{KataQMC}
G.~De~Filippis, V.~Cataudella, S.~Mishchenko, A.\, N.~Nagaosa, A.~Fierro, 
  A.~de~Candia, 
\newblock {\em Phys. Rev. Lett.}, {\bf  2015}, 114, 086601.

\bibitem{YaoYao-JCP12}
Y. Yao, W. Si, X. Hou,  C.-Q. Wu,
\newblock {\em J. Chem. Phys.}, {\bf 2012}, 136, 234106.

\bibitem{MayouPRL2000}
D. Mayou,
\newblock {\em Phys. Rev. Lett.}, {\bf  2000}, 85, 1290.

\bibitem{TramblyPRL2006}
a) G. Trambly~de Laissardi\`ere, J.-P. Julien,  D. Mayou,
\newblock {\em Phys. Rev. Lett.}, {\bf  2006}, 97, 026601; b)
D. Nguyen-Manh, D. Mayou, G. J. Morgan,  A. Pasturel,  {\em J. Phys. F: Met. Phys.} \textbf{1987}, 17, 999;
c) E. Belin,  D. Mayou, {\em Phys. Scripta} \textbf{1993}, T49, 356. 

\bibitem{MoixNJP2013} J. M. Moix, M. Khasin,  J. Cao, {\em New J. Phys.} 
{\bf 2013}, 15, 085010.

\bibitem{Packwood2015} D. M. Packwood, K. Oniwa, T. Jin,  N. Asao, 
{\em J. Chem. Phys.} {\bf 2015}, 142, 144503. 

\bibitem{DresselGrunerBook}
M. Dressel, G. Grüner,
\newblock {\em Electrodynamics of Solids - Optical Properties of Electron in
  Matter}.
\newblock Cambridge University Press, 2002.

\bibitem{ThoulessPRL1977}
D.~J. Thouless,
\newblock {\em Phys. Rev. Lett.}, {\bf  1977}, 39, 1167.

\bibitem{Mayougraphene13}
a) G. Trambly~de Laissardi\`ere,  Didier Mayou,
\newblock {\em Phys. Rev. Lett.}, {\bf  2013}, 111, 146601; b)
P. Darancet, V. Olevano, D. Mayou. {\em Phys. Rev. B} \textbf{2010}, 81, 155422; c)
S. Roche, D. Mayou, {\em Phys. Rev. B} \textbf{1999}, 60, 322; d)
D. Mayou, S. Khanna, {\em J. Phys. I},  \textbf{1995}, 5, 1199; e)
D. Mayou {\em Europhys. Lett.} \textbf{1988}, 6, 549.


\bibitem{Dong-highmobility-review-AdvMat13}
H. Dong, X. Fu, J. Liu, Z. Wang,  W. Hu,
\newblock {\em Adv. Mater.}, {\bf 2013}, 25, 6158.

\bibitem{BaoNcomms2014}
Y. Yuan, G. Giri, A.~L. Ayzner, A.~P. Zoombelt, S. C.~B.
  Mannsfeld, J. Chen, D. Nordlund, M.~F. Toney, J. Huang, 
  Z, Bao,
\newblock {\em Nat. Commun.}, {\bf  2014}, 5, 3005.

\bibitem{GarryCM2013}
K.~A. McGarry, W. Xie, C. Sutton, C. Risko, Y. Wu,
  V.~G. Young, J.-L. Br\'edas, C.~D. Frisbie,  C.~J.
  Douglas,
\newblock {\em Chem. Mater.}, {\bf 2013}, 25, 2254.

\bibitem{TullySurfaceHopping}
J.~C. Tully,
\newblock {\em J. Chem. Phys.}, {\bf 1990}, 93, 1061.

\bibitem{Akimov2014}
A.~V.~Akimov,  R.~Long, O.~V.~Prezhdo,
\newblock {\em J. Chem. Phys.}, {\bf  2014}, 140, 194107.

\bibitem{KataPRB2011}
V.~Cataudella, G.~De~Filippis,  C.~A. Perroni,
\newblock {\em Phys. Rev. B}, {\bf  2011}, 83, 165203.

\bibitem{SvistunovDIAGMCPRL1998}
N.~V. Prokof'ev, B.~V. Svistunov,
\newblock {\em Phys. Rev. Lett.}, {\bf  1998}, 81, 2514.

\bibitem{JarrellMaxEntPhysRep1996}
M. Jarrell,  J.E. Gubernatis,
\newblock {\em Phys. Rep.}, {\bf 1996}, 269, 133.

\bibitem{KataSSHPRL2010}
D.~J.~J. Marchand, G.~De~Filippis, V.~Cataudella, M.~Berciu, N.~Nagaosa, N.~V.
  Prokof'ev, A.~S. Mishchenko,  P.~C.~E. Stamp,
\newblock {\em Phys. Rev. Lett.}, {\bf  2010}, 105,266605.

\bibitem{Okamoto1}
R.~Uchida, H.~Yada, M.~Makino, Y.~Matsui, K.~Miwa, T.~Uemura, J.~Takeya, 
  H.~Okamoto,
\newblock {\em Appl. Phys. Lett.}, {\bf 2013}, 102, 093301.

\bibitem{GogolinJETP75}
V.~I. Mel'nikov, A. A.  Gogolin, E. I.  Rashba,
\newblock {\em Sov. Phys. JETP}, {\bf 1975},  42, 168.

\bibitem{RashbaGogolinMelnikov76}
E.I. Rashba, A.A. Gogolin,  V.I. Mel'nikov,
\newblock In  {\em
  Organic Conductors and Semiconductors}, volume~65 of {\em Lecture Notes in
  Physics}, (Eds. L.~Pal, G.~Gruner, A.~Janossy, and J.~Solyom), 
  Springer Berlin Heidelberg, {\bf 1977}, pp. 265--280.

\bibitem{GirvinPRL1979}
M.~Jonson,  S.~M. Girvin.
\newblock {\em Phys. Rev. Lett.}, {\bf  1979}, 43, 1447.

\bibitem{MadhukarCohenPRL1977}
A. Madhukar,  M.~H. Cohen, 
\newblock {\em Phys. Rev. Lett.}, {\bf  1977}, 38, 85.

\bibitem{GogolinMelnikovRashbaJETP77}
A.~A.~Gogolin,  V.~I.~Mel'nikov, E.~I. Rashba,
\newblock {\em Sov. Phys. JETP}, {\bf 1977},  45, 330.

\bibitem{GogolinPhysRep81}
a) A.A. Gogolin,
\newblock {\em Phys. Rep.}, {\bf 1982}, 86, 1;
b) A.A. Gogolin,
\newblock {\em Phys. Rep.}, {\bf 1988}, 166, 269.

\bibitem{GogolinZimanyiSSC83}
A.A. Gogolin,  G.T. Zimanyi,
\newblock {\em Solid State Commun.}, {\bf 1983}, 46, 469.

\bibitem{GogolinZimanyiSSC84}
A.A. Gogolin, G.T. Zimanyi,
\newblock {\em Solid State Commun.}, {\bf 1984}, 50, 791.

\bibitem{ShanteJPC1978}
V.~K.~S. Shante, 
\newblock {\em J. Phys. C: Solid State Physics}, {\bf 1978},
  11, 2561.

\bibitem{TakenakaPRL05}
K.~Takenaka, M.~Tamura, N.~Tajima, H.~Takagi, J.~Nohara,  S.~Sugai,
\newblock {\em Phys. Rev. Lett.}, {\bf  2005}, 95, 227801.

\bibitem{Hashimoto} K. Hashimoto, S. C. Zhan, R. Kobayashi, S. Iguchi, N. Yoneyama, T. Moriwaki, 
Y. Ikemoto, T. Sasaki, {\em Phys. Rev. B}, \textbf{2014}, 89, 085107.

\bibitem{Mahmoudian} S. Mahmoudian,  L. Rademaker, A. Ralko, S. Fratini, V. Dobrosavljevi\'c, preprint arXiv:1412.4441.

\bibitem{SeoJPSJ2006}
H. Seo, J. Merino, H. Yoshioka,  M. Ogata,
\newblock {\em J. Phys. Soc. Jpn.}, {\bf 2006},
  75, 051009.

\bibitem{CanoPRL2010}
L.~Cano-Cort\'es, J.~Merino,  S.~Fratini,
\newblock {\em Phys. Rev. Lett.}, {\bf  2010}, 105, 036405.

\bibitem{Graja}
A. Graja,
\newblock {\em Low-Dimensional Organic Conductors},
\newblock World Scientific, {\bf 1992}.

\bibitem{Dressel} M. Dressel, N. Drichko, {\em Chem. Rev. }, {\bf 2004}, 104, 5689.

\bibitem{XiePRL2014}
W. Xie, S. Wang, X. Zhang, C.~Leighton,  C.~D. Frisbie,
\newblock {\em Phys. Rev. Lett.}, {\bf  2014}, 113, 246602.

\bibitem{MartelPRL2008}
T.~Kampfrath, K.~von Volkmann, C.~M. Aguirre, P.~Desjardins, R.~Martel,
  M.~Krenz, C.~Frischkorn, M.~Wolf,  L.~Perfetti,
\newblock {\em Phys. Rev. Lett.}, {\bf  2008}, 101, 267403.

\bibitem{IchidaSSC2011}
M. Ichida, S. Saito, T. Nakano, Y.~Feng, Y. Miyata, K.
  Yanagi, H. Kataura,  H. Ando,
\newblock {\em Solid State Commun.}, {\bf 2011}, 151, 1696.

\bibitem{UlbrichtRMP2011}
R. Ulbricht, E. Hendry, J. Shan, T.~F. Heinz,  M. Bonn,
\newblock {\em Rev. Mod. Phys.}, {\bf  2011}, 83, 543.

\bibitem{Engel-Nature07}
G.~S. Engel, T.~R. Calhoun, E.~L. Read, T.-K. Ahn, T.
  Mancal, Y.-C. Cheng, R.~E. Blankenship,  G.~R. Fleming,
\newblock {\em Nature}, {\bf  2007}, 446, 782.

\bibitem{Mohseni-Rebentrost-JCP08}
M. Mohseni, P. Rebentrost, S. Lloyd, A. Aspuru-Guzik,
\newblock {\em J. Chem. Phys.}, {\bf 2008}, 129, 174106.

\bibitem{Rebentrost-NJP09}
P. Rebentrost, M. Mohseni, I. Kassal, S. Lloyd, A.
  Aspuru-Guzik, 
\newblock {\em New J. Phys.}, {\bf 2009}, 11, 033003.

\end{thebibliography}

\end{document}